\documentclass[5p,twocolumn]{elsarticle}

\usepackage[caption = false]{subfig}
\usepackage{graphicx}
\usepackage{dcolumn} 
\usepackage{bm}
\usepackage{amssymb}
\usepackage{verbatim}
\usepackage{booktabs}
\usepackage{amsmath}
\usepackage{color}

\newcommand{\Plossr}{P_{\rm loss\,r}^{\rm azi\,B}}
\newcommand{\Plossphi}{P_{\rm loss\,\phi}^{\rm rad\,B}}

\begin{document}

\begin{frontmatter}

\title{Concept of multiple-cell cavity for axion dark matter search}

\author[kaist]{Junu Jeong}
\author[ibs]{SungWoo Youn\corref{swyoun}}\ead{swyoun@ibs.re.kr}
\author[kaist]{Saebyeok Ahn}
\author[kyunghee,ibs]{Jihn E. Kim}
\author[kaist,ibs]{Yannis K. Semertzidis}
\address[kaist]{Department of Physics, Korea Advanced Institute of Science and Technology (KAIST), Daejeon 34141, Republic of Korea}
\address[ibs]{Center for Axion and Precision Physics Research, Institute for Basic Science, Daejeon 34047, Republic of Korea}
\address[kyunghee]{Department of Physics, Kyung Hee University, Seoul 02447, Republic of Korea}
\cortext[swyoun]{Corresponding author. Tel.: +82 42 350 7324}

\begin{abstract}
In cavity-based axion dark matter search experiments exploring high mass regions, multiple-cavity design is under consideration as a method to increase the detection volume within a given magnet bore. 
We introduce a new idea, referred to as a multiple-cell cavity, which provides various benefits including a larger detection volume, simpler experimental setup, and easier phase-matching mechanism.
We present the characteristics of this concept and demonstrate the experimental feasibility with an example of a double-cell cavity.
\end{abstract}


\end{frontmatter}

\section{Introduction}
The axion is a hypothetical elementary particle postulated to solve the strong CP problem in quantum chromodynamics of particle physics~\cite{bib:PQWW}.
With its mass falling in a specific range, the main cosmological interest of the axion at present is its role as a candidate for cold dark matter (CDM)~\cite{bib:CDM}, which would account for 27\% of the energy of our Universe~\cite{bib:universe_comp}.
If it is CDM, the axion must be an ``invisible (very light)" particle of a type of KSVZ~\cite{bib:KSVZ}, DFSZ~\cite{bib:DFSZ}, or their combinations.

A conventional axion dark matter search experiment using microwave resonant cavities adopts the methodological concept proposed by P. Skivie based on the Primakoff effect~\cite{bib:Primakoff}, in which axions are converted into radio-frequency (RF) photons in a strong magnetic field, which are in turn resonated in microwave cavities immersed in the field~\cite{bib:Sikivie}.
The axion-to-photon conversion power is given by 
\begin{equation}
P_{a\rightarrow\gamma\gamma} = g_{a\gamma\gamma}^2 \frac{\rho_a}{m_a}B_0^2VC\,{\rm{min}}(Q_L,Q_a),
\label{eq:power}
\end{equation}
where $g_{a\gamma\gamma}$ is the axion-photon coupling constant, $\rho_a$ is the axion local density, $m_a$ is the axion mass, $B_0$ is the external magnetic field, $V$ is the cavity volume, and $Q_L$ and $Q_a$ are the quality factors of the loaded cavity and of the axion, respectively~\cite{bib:detection_rate}.
The mode-dependent form factor $C$ is defined as
\begin{equation}
C=\frac{\left| \int_V \mathbf{E_c}\cdot \mathbf{B_0} d^3x\right|^2}{\int_V \epsilon(x) |\mathbf{E_c}|^2 d^3x \int_V |\mathbf{B_0}|^2 d^3x},
\label{eq:form_factor}
\end{equation}
where $\mathbf{B_0}$ is the external magnetic field, $\mathbf{E_c}$ is the electric field of the cavity resonant mode under consideration, and $\epsilon(x)$ is the dielectric constant inside the cavity volume.
For cylindrical cavities in a solenoidal magnetic field, the TM$_{010}$ mode is commonly considered because it gives the largest form factor.
Since the mass of the axion is a priori unknown, a cavity must be tunable to search for a signal over the frequency range determined by the cavity design.
It is important to scan a large frequency range as fast as possible with a given experimental sensitivity.
A relevant quantity to this is the scan rate whose maximum value is obtained as
\begin{equation}
\frac{d f}{d t} = \left(\frac{1}{\rm{SNR}}\right)^2\left(\frac{P_{a\rightarrow\gamma\gamma}}{k_BT_{\rm{sys}}}\right)^2  \frac{Q_a}{Q_L},
\label{eq:snr}
\end{equation}
where SNR is the signal-to-noise ratio, $k_B$ is the Boltzmann constant, and $T_{\rm{sys}}$ is the total noise temperature of the system.

Exploring higher frequency regions in axion search experiments using microwave cavity detectors requires a smaller size of cavity because the TM$_{010}$ resonant frequency is inversely proportional to the cavity radius. 
An intuitive way to make effective use of a given magnet volume, and thereby to increase the experimental sensitivity, is to bundle an array of identical cavities together and combine their individual outputs ensuring phase matching of the coherent axion signal~\cite{bib:multicavity}.
This conventional approach has been experimentally attempted using a quadruple-cavity system~\cite{bib:ADMX_multicav} but its methodological advantage was not well addressed, and thereby led to a further in-depth study~\cite{bib:multiple_cavity}.

\begin{figure}[h]
\centering
\subfloat[]{\includegraphics[width=0.125\textwidth]{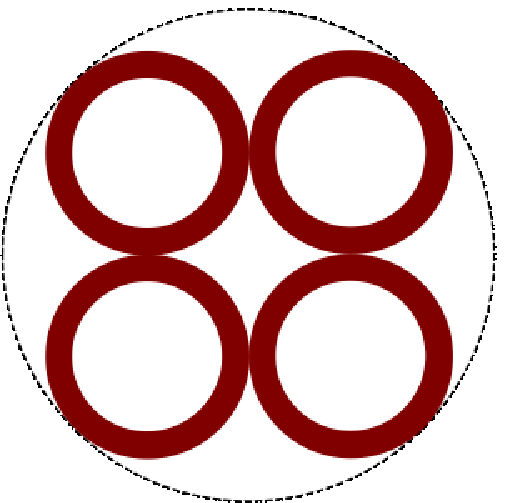}}
\hspace{0.03\textwidth}
\subfloat[]{\includegraphics[width=0.125\textwidth]{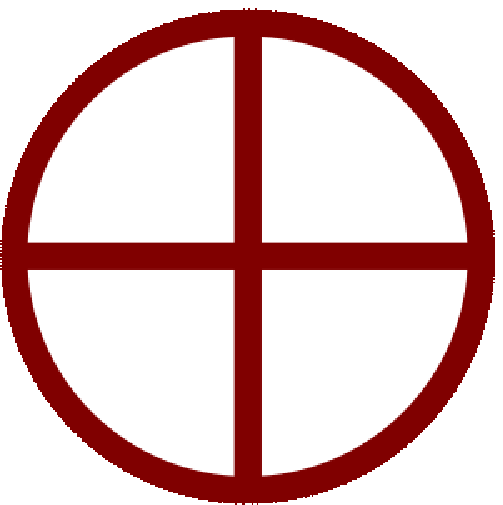}}
\hspace{0.03\textwidth}
\subfloat[]{\includegraphics[width=0.125\textwidth]{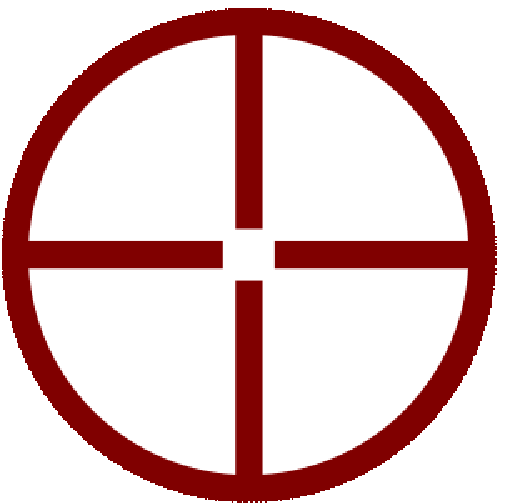}}
\caption{Various designs of multiple-detector system: (a) multiple-cavity; (b) multiple-cell cavity; and (c) multiple-cell cavity with a hollow gap in the middle. 
The dashed line represents the boundary of the magnet bore.}
\label{fig:config}
\end{figure}

The multiple-cavity design, however, is still inefficient in volume usage for a given magnet, mainly due to unused volume and cavity wall thickness, as can be seen in Fig.~\ref{fig:config}(a).
One alternative design is a single cylindrical cavity, fitting into the magnet bore, split by metal partitions placed at equidistant intervals to make multiple identical cells, as shown in Fig.~\ref{fig:config}(b).
This concept, initially discussed in Ref.~\cite{bib:multicell}, provides a more effective way to increase the detection volume while relying on the same frequency tuning mechanism as that of multiple-cavity systems.
The resonant frequency increases with the cell multiplicity.
Furthermore, an innovative idea of introducing a narrow hollow gap at the center of the cavity, as seen in Fig.~\ref{fig:config}(c), provides critical advantages.
In this design, all cells are spatially connected among others, which allows a single RF coupler to extract the signal out of the entire cavity volume.
This simplifies the readout chain by not only reducing the number of RF antennae but also eliminating the necessity of a power combiner, both of which could be bottlenecks for multiple-cavity systems especially when the cavity multiplicity is large.
We refer to this cavity concept as a pizza cylinder cavity.

Using the COMSOL Multiphysics$\textsuperscript{\textregistered}$ software~\cite{bib:comsol}, this multiple-cell design is compared with the conventional multiple-cavity design in terms of various experimental quantities. 
Assuming a magnet bore diameter of 100\,mm, a cavity height of 200\,mm, and a wall thickness of 5 mm, various quantities for a quadruple-cavity system and a quadruple-cell cavity are summarized in Table~\ref{tab:comp}.
For a more reasonable comparison, a sextuple-cell cavity whose resonant frequency is similar to that of the quadruple-cavity system is also considered.
It is remarkable that, mainly due to the volume increase, the conversion power and scan rate are significantly improved.
\begin{table}[h]
\caption{\label{tab:comp} Comparison of experimental quantities among various designs of multiple-detector system. 
$V$ is the total detection volume, $Q$ is the unloaded quality factor at room temperature, and the conversion power and scan rate are values relative to the quadruple-cavity detector.
The externally applied magnetic field is assumed to be homogenous within the magnet bore.}
\begin{tabular*}{0.475\textwidth}{@{\extracolsep{\fill}}cccc}
 \toprule
 & quad-cavity & quad-cell & sext-cell \\
 \midrule
$V$ [L] & 0.62 & 1.08 & 1.02 \\
$f_{\rm{TM_{010}}}$ [GHz]& 7.30 & 5.89 & 7.60 \\
$Q$ & 19,150 & 19,100 & 16,910 \\
$C$ & 0.69 & 0.65 & 0.63 \\
$P_{a\rightarrow\gamma\gamma}$ & 1.00 & 1.65 & 1.32 \\
$d^{}f/d^{}t$ & 1.00 & 2.72 & 1.98 \\
\bottomrule
\end{tabular*}
\label{tab:comp}
\end{table}

\section{Analytical EM solution for multiple-cell cavity}
\subsection{Field solution for perfect electric conductor}
\begin{figure}[h]
\centering
\subfloat[]{\includegraphics[width=0.2\textwidth]{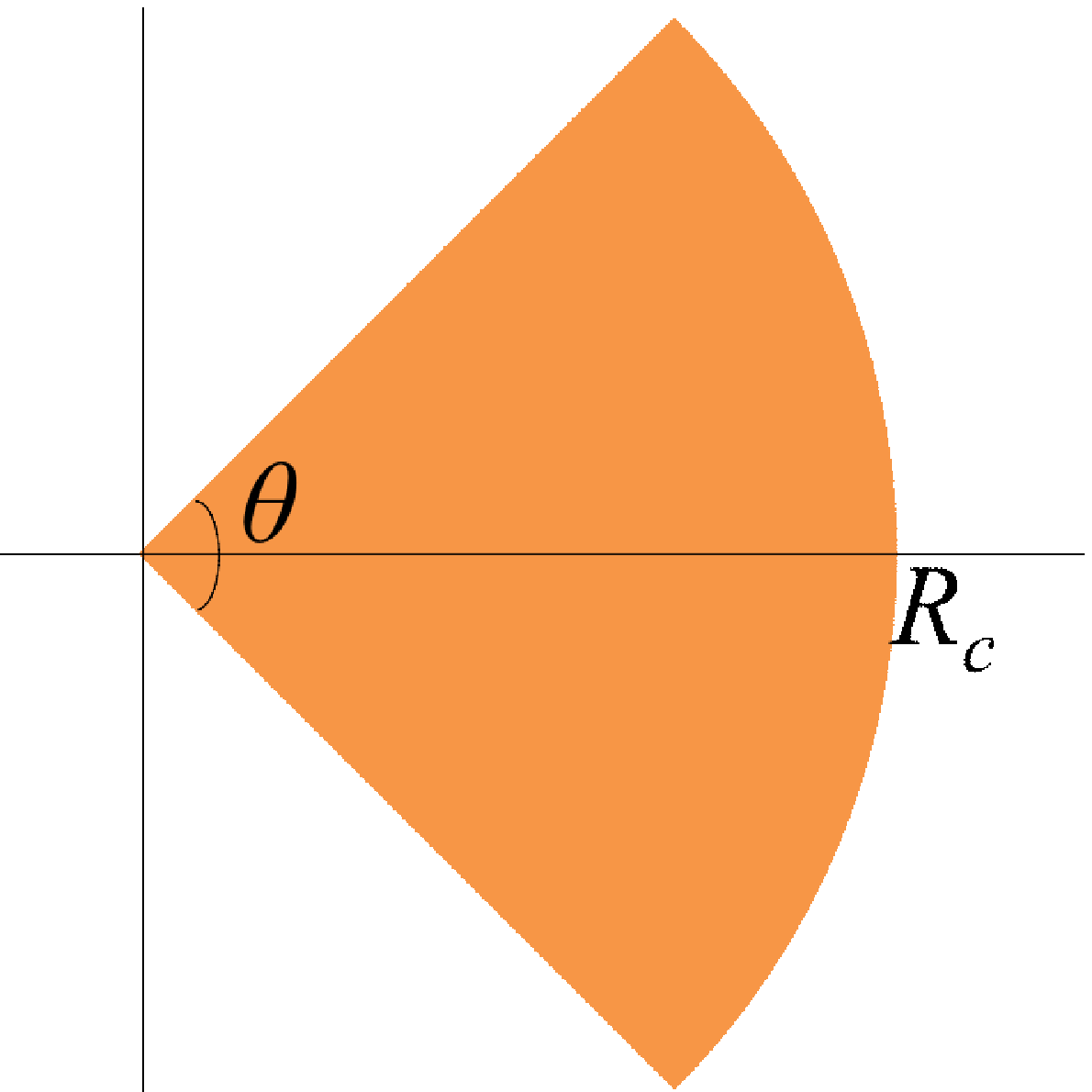}}
\hspace{0.02\textwidth}
\subfloat[]{\includegraphics[width=0.2\textwidth]{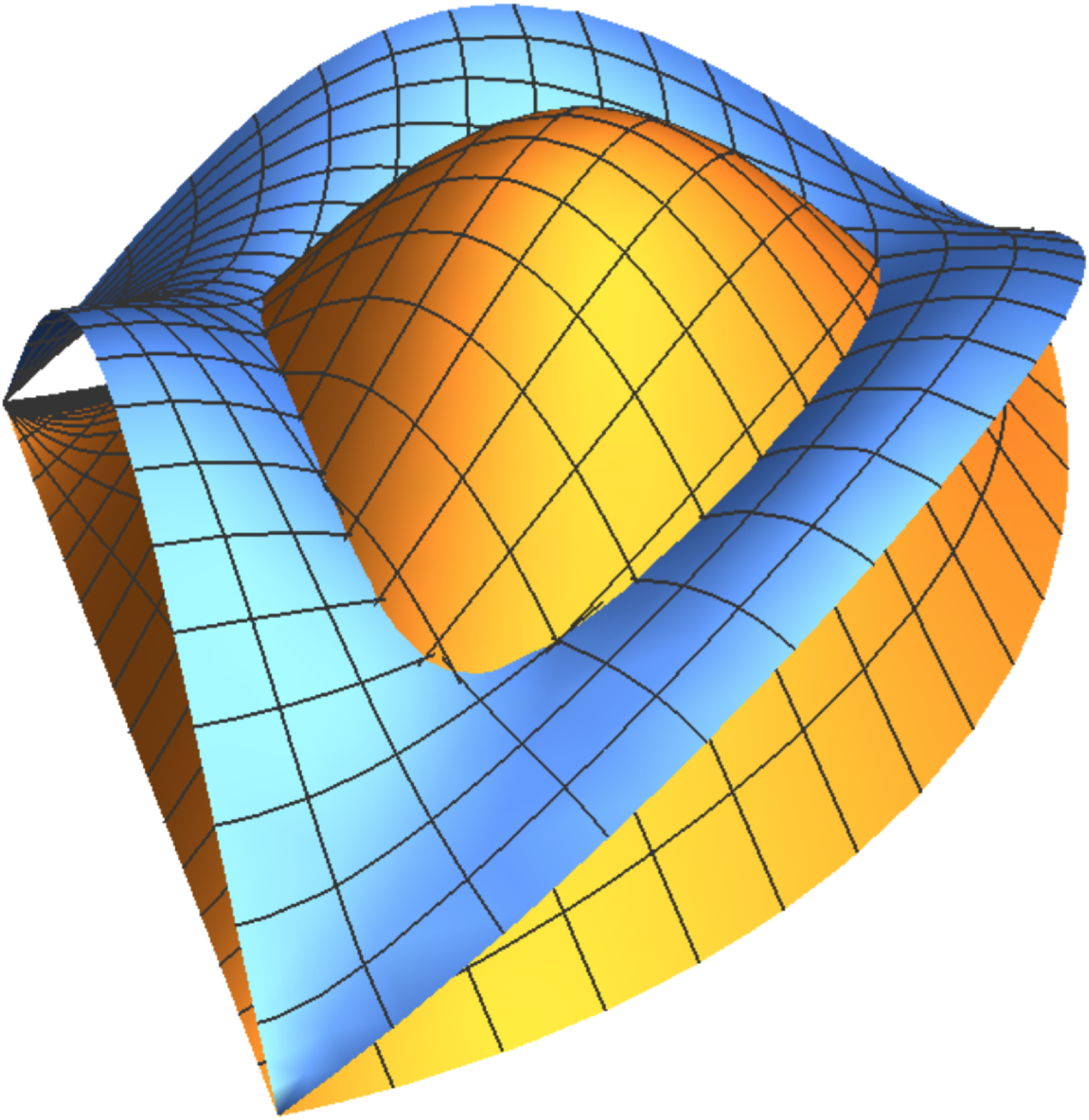}}
\caption{(Color online.) (a) Cross-sectional view of a cell of a PEC with $\theta=\pi/4$ and (b) 3D view of the corresponding electric (orange) and magnetic (blue) field distributions for the TM$_{010}$-like mode.}
\label{fig:solution}
\end{figure}

With a static external magnetic field $\mathbf{B_0}=B_0\hat{z}$, the axion field $a$ modifies Maxwell equations as in~\cite{bib:electrodynamics}
\begin{equation}
\begin{split}
\nabla\cdot\mathbf{E} &= -g\nabla a\cdot \mathbf{B} \approx 0 \\
\nabla\times\mathbf{E} &= - \partial \mathbf{B} /\partial t \\
\nabla\cdot\mathbf{B} &= 0\\
\nabla\times\mathbf{B} &= \partial \mathbf{E} /\partial t + g(\partial a /\partial t \mathbf{B_0}+\nabla a \times\mathbf{E}) \\
&\approx  \partial \mathbf{E} /\partial t + g\partial a /\partial t \mathbf{B_0},
\end{split}
\label{eq:Maxwell}
\end{equation}
where $g\equiv g_{a\gamma\gamma}$ and $\nabla a\approx 0$ are attributed to the large de Broglie wavelength of the axion field~\cite{bib:multiple_cavity}.
The Maxwell-Ampere equation, Eq.~(\ref{eq:Maxwell}), is expressed in terms of the vector potential $\mathbf{A}$ with a choice of the Coulomb gauge, $\nabla\cdot\mathbf{A} = 0$, of
\begin{equation}
\nabla\times\nabla\times\mathbf{A} + \partial^2 \mathbf{A} /\partial t^2 = g\partial a /\partial t \mathbf{B_0}.
\label{eq:4thME}
\end{equation}
For a cylindrical cavity system, since the EM fields of the TM mode under our consideration form a symmetry in the $z$ direction, the radial and azimuthal components of the vector potential are independent from the vertical component, which leads to:
\begin{equation}
A_r = 0 \textrm{\;and\;} A_{\phi}=0.
\end{equation}
With the oscillating axion field $a=a_0e^{-i\omega t}$, Eq.~(\ref{eq:4thME}) gives
\begin{equation}
\frac{1}{r}\frac{\partial}{\partial r}\left( r\frac{\partial A_{z}}{\partial r}\right)+\frac{1}{r^{2}}\frac{\partial^{2} A_{z}}{\partial \phi^{2}}+\frac{\partial^{2} A_{z}}{\partial z^{2}}+\omega^{2}A_{z}=-i\omega gB_{0}a.
\label{eq:Az}
\end{equation}
Using an ansatz $A_z= R(r)\Phi(\phi) + \Omega(\omega)$, we obtain the general solutions to Eq.~(\ref{eq:Az}) as
\begin{equation}
\begin{split}
\Omega(\omega) &= -\frac{igB_{0}a}{\omega}\\
\Phi(\phi)&= \{e^{\pm im\phi}\}\\
R(r)&= \{J_{m}(r\omega), Y_{m}(r\omega)\},
\end{split}
\label{eq:gen_sol}
\end{equation}
where $J_{m} (Y_{m})$ is the Bessel function of the first (second) kind for integer or positive $m$.
The curly brackets in Eq.~(\ref{eq:gen_sol}) represent linear combinations of the elements inside the brackets.
By requiring $A_z(\phi)=A_z(-\phi)$ and $R(0)=\mathrm{finite}$, we obtain the solution for the vector potential,
\begin{equation}
A_{z}(r,\phi)=A_{0}J_{m}(r\omega)\cos{m\phi}-\frac{igB_{0}a}{\omega},
\end{equation}
from which the EM field solutions are obtained:
\begin{equation}
\begin{split}
E_{z}(r,\phi)&=i\omega A_{0}J_{m}(r\omega)\cos{m\phi}+gB_{0}a\\
B_{r}(r,\phi)&=-\frac{A_{0}}{r}J_{m}(r\omega)m\sin{m\phi}\\
B_{\phi}(r,\phi)&=-\frac{A_{0}\omega}{2}(J_{m-1}(r\omega)-J_{m+1}(r\omega))\cos{m\phi}.
\end{split}
\label{eq:solution}
\end{equation}
If we define the enhancement factor $Q_J\equiv | \frac{i\omega A_0}{gB_0a}|$, the source term $gB_{0}a$ can be ignored on resonance since $Q_J \gg 1$.
Then the boundary conditions for a perfect electric conducting (PEC) cell as shown in Fig~\ref{fig:solution}(a), $E_{z}(r=R_c,\phi)=0$ and $E_{z}(r,|\phi|=\theta/2)=0$, yield
\begin{equation}
\omega=\frac{\chi_{mn}}{R}\;{\rm{and}}\;m=\frac{\pi}{\theta},
\end{equation}
where $\chi_{mn}$ is the $n$-th root of the Bessel function of order $m$. 
The EM field distributions for the lowest TM mode of a PEC cell with $\theta=\pi/4$ are shown in Fig.~\ref{fig:solution}(b).

\subsection{Quality factor for normal conductor}
The quality factor of a multiple-cell cavity of a normal conductor can be calculated by an approximation to PEC:
\begin{equation}
Q=\omega\frac{U}{P_{\rm{loss}}}
\label{eq:Q}
\end{equation}
where $U$ is the total stored energy in the cells, which is given by
\begin{equation}
U=\frac{\epsilon}{2}\int_V |\mathbf{E}|^2 dV.
\end{equation}
The power loss at the conductor walls is calculated using 
\begin{equation}
P_{\rm{loss}}=\frac{1}{2\sigma\mu^2\delta} \int_{A}dA|\mathbf{n}\times\mathbf{B}|^{2},
\end{equation}
where $\sigma$ is the conductivity, $\mu$ is the permeability, and $\delta$ is the skin depth.
Table~\ref{tab:Q} shows a good agreement in the quality factor between the analytical calculation from Eq.~(\ref{eq:Q}) and the numerical approximation from COMSOL simulation for various cell multiplicities.
\begin{table}[h!]
\caption{Comparison of quality factors from analytical solution and numerical simulation for a multiple-cell cavity with no end caps.
The cavity is assumed to be made of nominal copper with $R_c=25$\,mm and $\sigma=5.96\times 10^{7}\,$S/m at room temperature.
The superscript $^{\rm ax}$ indicates the numerical calculation in the presence of the axion source.
The numbers in parentheses are deviations from the analytical values in percentage.}
\begin{tabular*}{0.475\textwidth}{@{\extracolsep{\fill}}ccccc}
\toprule
\# cells & $\theta$  & Analytical  & Numerical & Numerical$^{\rm ax}$ \\
\midrule
2	&$\pi$	& 17,794	& 17,819 (0.1)	& 17,836 (0.2)\\ 
4	& $\pi/2$	& 16,151	& 16,151 (0.0)	& 16,335 (1.1)\\
6	& $\pi/3$	& 14,320	& 14,310 (0.1)	& 14,325 (0.0)\\
8	& $\pi/4$	& 12,907	& 12,851 (0.4)	& 12,930 (0.2)\\
\bottomrule
\label{tab:Q}
\end{tabular*}
\end{table}

It would be useful to evaluate the relative effect of the sides of a cell on the quality factor.
We separate the total power loss into two components: $\Plossphi$ by the cell sides owing to the radial magnetic field and $\Plossr$ by the cell arc owing to the azimuthal magnetic field, such that
\begin{equation}
\begin{split}
P_{\rm{loss}} &= \Plossphi +\Plossr \\
&= \frac{1}{2\sigma\mu^2\delta} \left[ 2\int_{0}^{R_c} |\mathbf{B}_r|_{\phi=\theta/2}^2 dr + R\int_{-\theta/2}^{\theta/2} |\mathbf{B}_{\phi}|_{r=R_c}^2 d\phi \right].
\end{split}
\end{equation}
By plugging Eq. (\ref{eq:solution}) into Eq. (\ref{eq:P_ratio}), we find that the ratio $\Plossphi$/$\Plossr$ depends only on the sector angle $\theta$, i.e., cell multiplicity.
\begin{equation}
\frac{\Plossphi}{\Plossr} = \frac{4n^3}{n^2-1} \frac{J_{n/2+1}(\chi)^2}{\pi\left(J_{n/2} (\chi)-J_{n/2+1}(\chi)\right)^2},
\label{eq:P_ratio}
\end{equation}
where $\chi$ is the first root of $J_{n/2}(r)$.
The tendency of various quantities to vary with the cell multiplicity is summarized in Table~\ref{tab:P_loss}.

\begin{table}[h!]
\caption{Dependence of the TM$_{010}$-like resonant frequency, quality factor, and power loss ratio on the cell multiplicity, equivalently on the sector angle with the same geometry, listed in Table \ref{tab:Q}.
The numbers are values relative to those of the cavity with zero cell, which corresponds to the conventional cylindrical cavity.}
\begin{tabular*}{0.475\textwidth}{@{\extracolsep{\fill}}ccccc}
\toprule
\# cells & $\theta$  & $\omega/\omega_0$  & $Q/Q_0$ & $\Plossphi/\Plossr$ \\
\midrule
0	&no wall	& 1.00	& 1.00	& 0.00\\ 
2	&$\pi$	& 1.59	& 0.68	& 0.84\\ 
4	& $\pi/2$	& 2.14	& 0.62	& 1.36\\
6	& $\pi/3$	& 2.65	& 0.55	& 1.96\\
8	& $\pi/4$	& 3.16	& 0.50	& 2.59\\
\bottomrule
\label{tab:P_loss}
\end{tabular*}
\end{table}

\section{Characteristics of multiple-cell design}

\subsection{Hollow gap and frequency degeneracy breaking}\label{subsec:degeneracy}
\begin{figure}[h]
\includegraphics[width=0.115\textwidth]{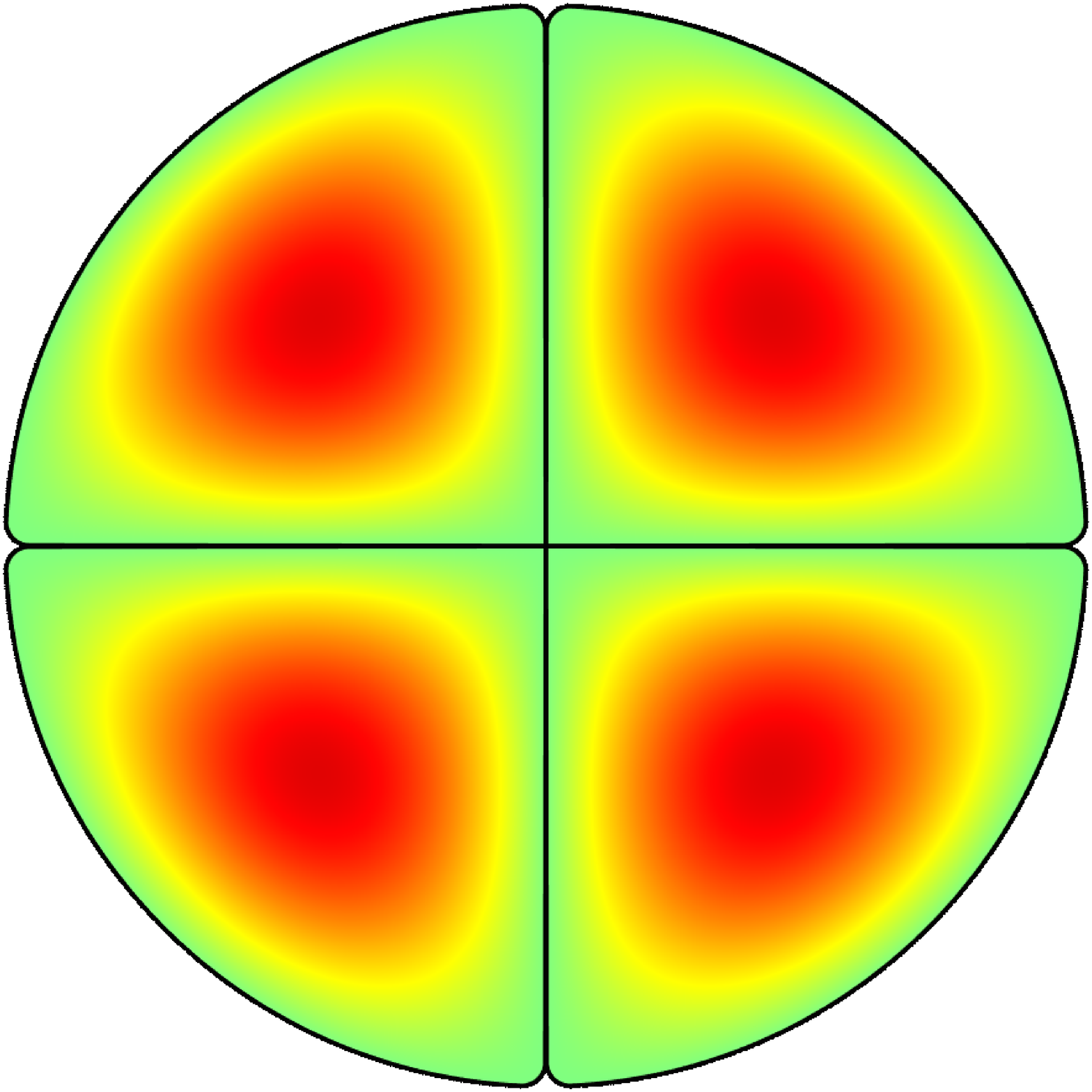}
\includegraphics[width=0.115\textwidth]{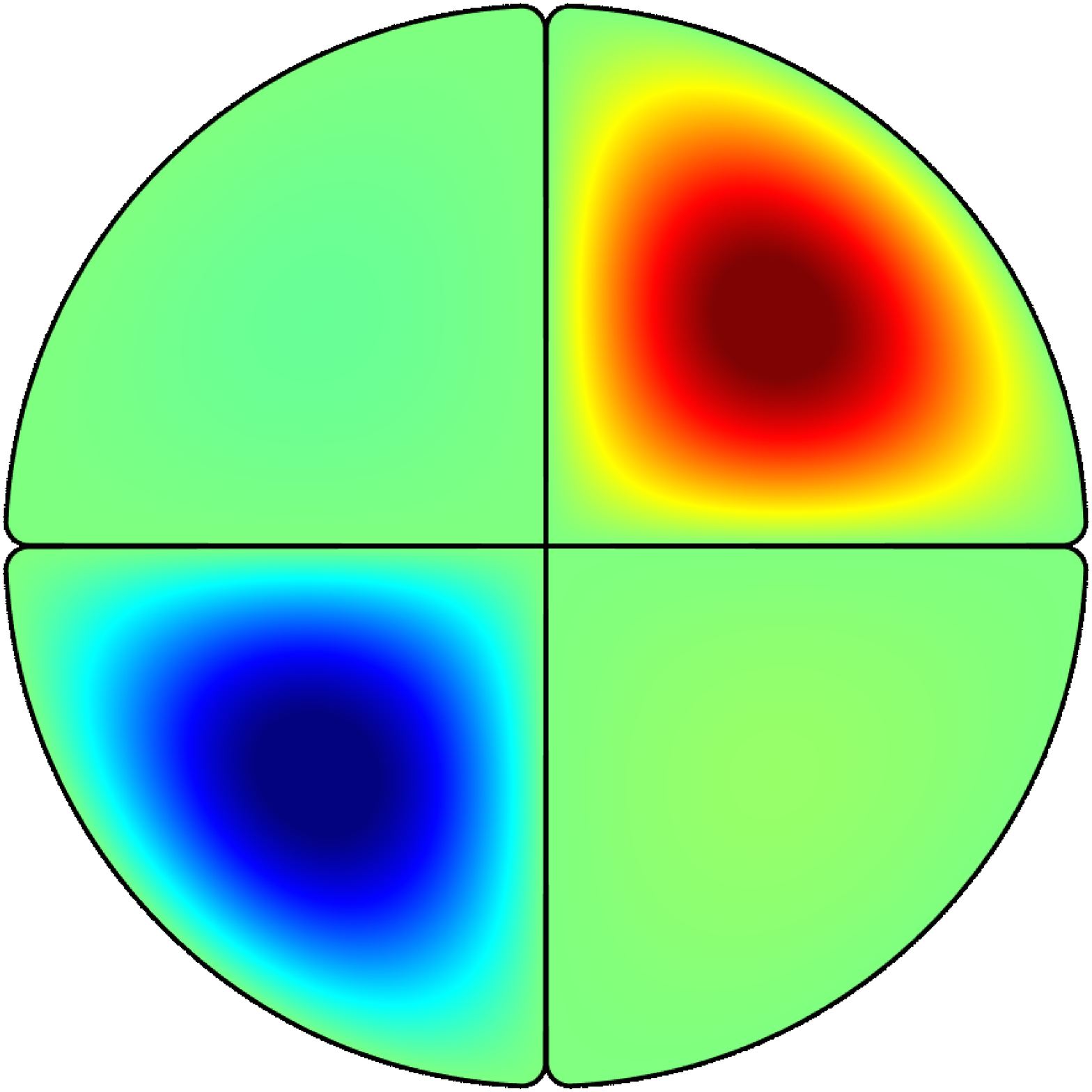}
\includegraphics[width=0.115\textwidth]{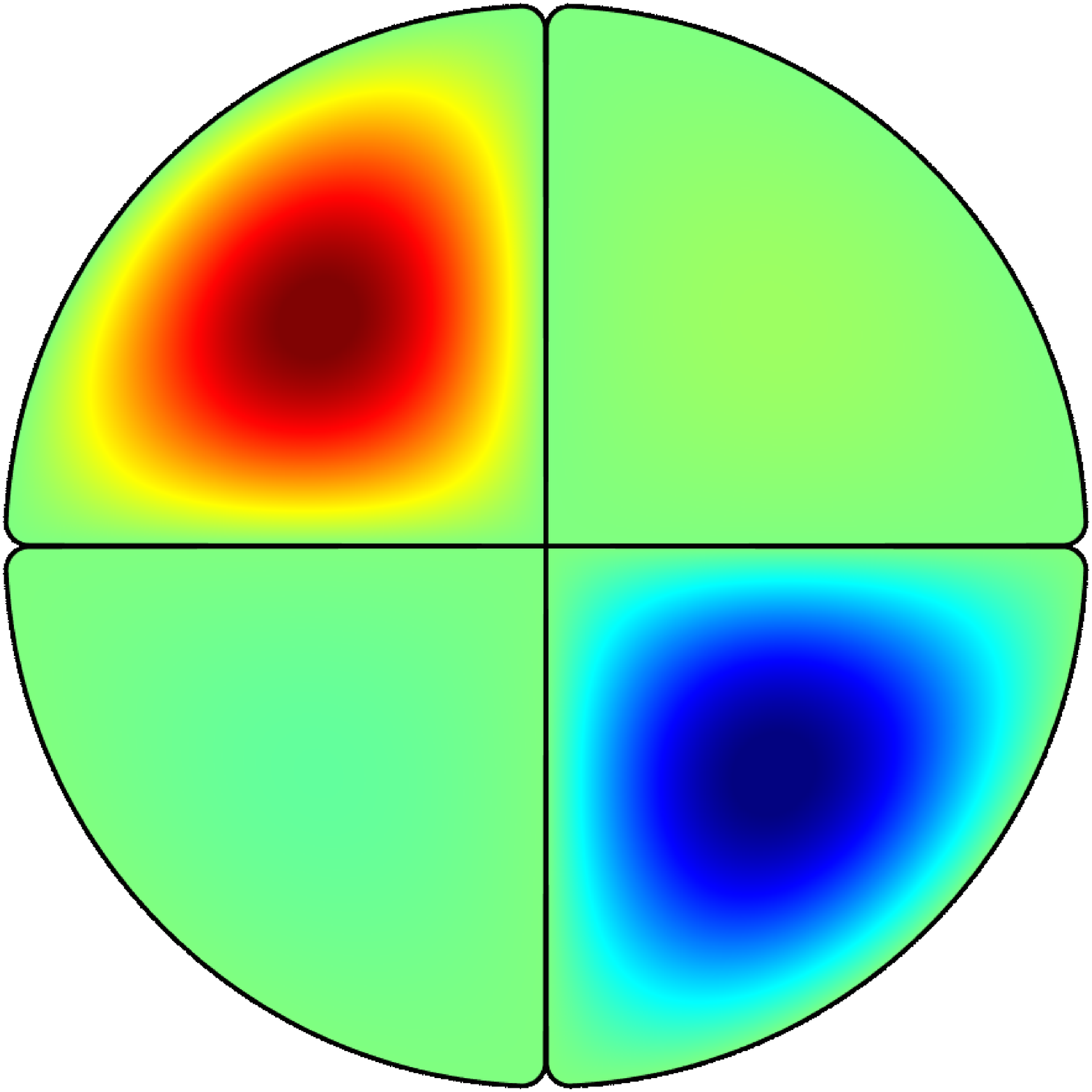}
\includegraphics[width=0.115\textwidth]{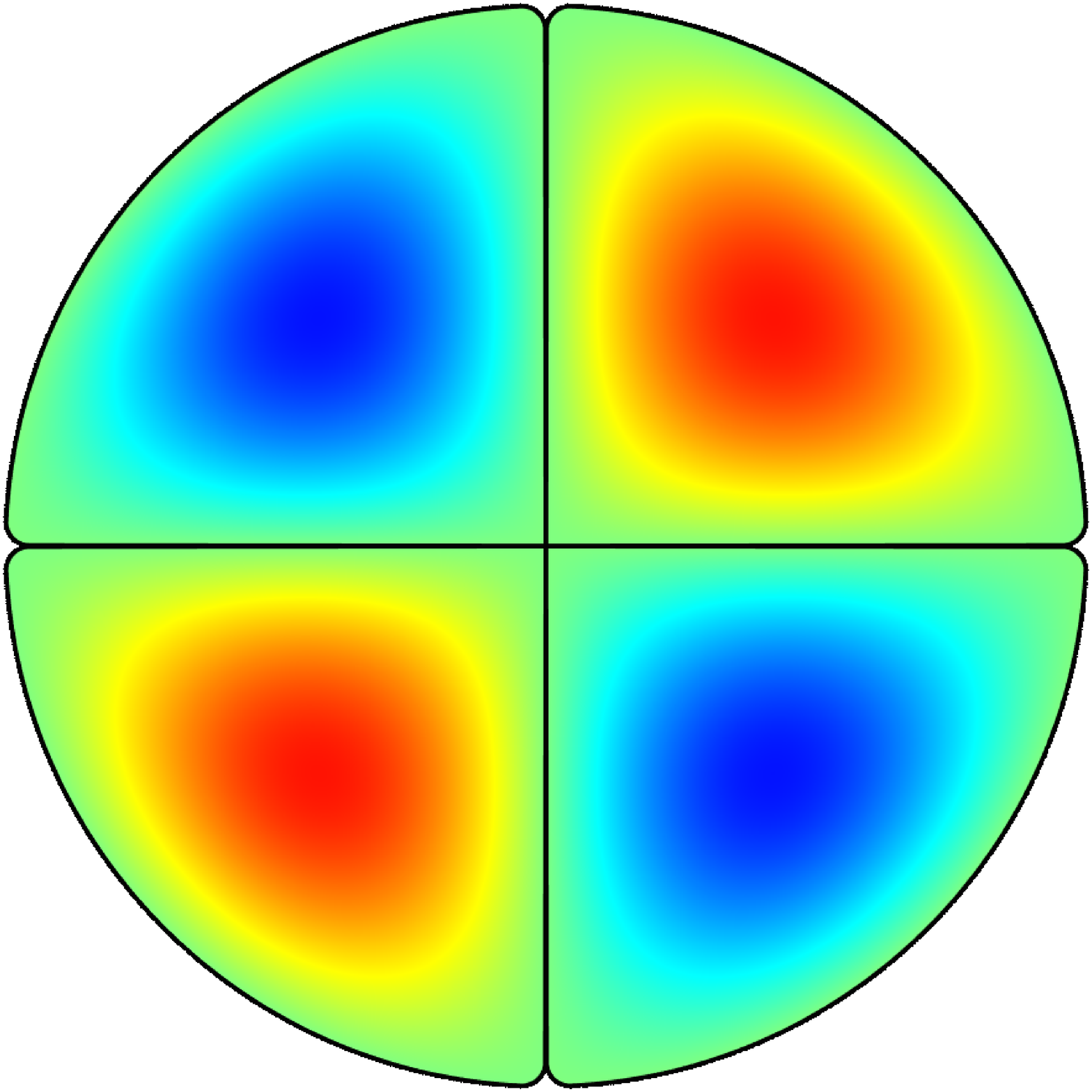}
\caption{(Color online.) Cross-sectional view of the electric field distribution of different modes for a quadruple-cell cavity: TM$_{010}$-like mode (left), two TM$_{110}$-like modes depending on polarization (middle), and TM$_{210}$-like mode (right).}
\label{fig:degenaracy}
\end{figure}

In the multiple-cell design, however, multiple modes can degenerate at the same frequency. 
For a quadruple-cell cavity, for instance, there are four degenerate modes possible depending on the overall electromagnetic (EM) field configuration, as shown in Fig.~\ref{fig:degenaracy}.
The mode with all individual EM fields in phase is referred to as the TM$_{010}$-like mode, while the one with the EM fields alternately 180 degree out of phase (counter)clockwise is referred to as the TM$_{210}$-like mode.
There are two additional modes in between, referred to as the TM$_{110}$-like modes, depending on polarization.
We find, based on the simulation study, that the frequency degeneracy becomes broken by the hollow gap in the middle of the cavity and that the degree of the degeneracy breaking, defined as the frequency separation between the first two lowest modes, depends on the size of the gap, i.e., the larger the hollow is, the larger is the separation; the system eventually approaches to a single cavity.
It is also noted that the lowest frequency corresponds to the TM$_{010}$-like mode, which is of our interest, regardless of the cell multiplicity.
Note that all higher modes are not important since the EM fields are out of phase,  and  the form factor,  viz. Eq. (\ref{eq:form_factor}), is always zero with an external static magnetic field. Thus, we consider only the lowest TM$_{010}$-like mode.

\subsection{Electric field distribution and phase-matching}\label{subsec:phase-matching}
\begin{figure}[h]
\centering
\includegraphics[width=0.14\textwidth]{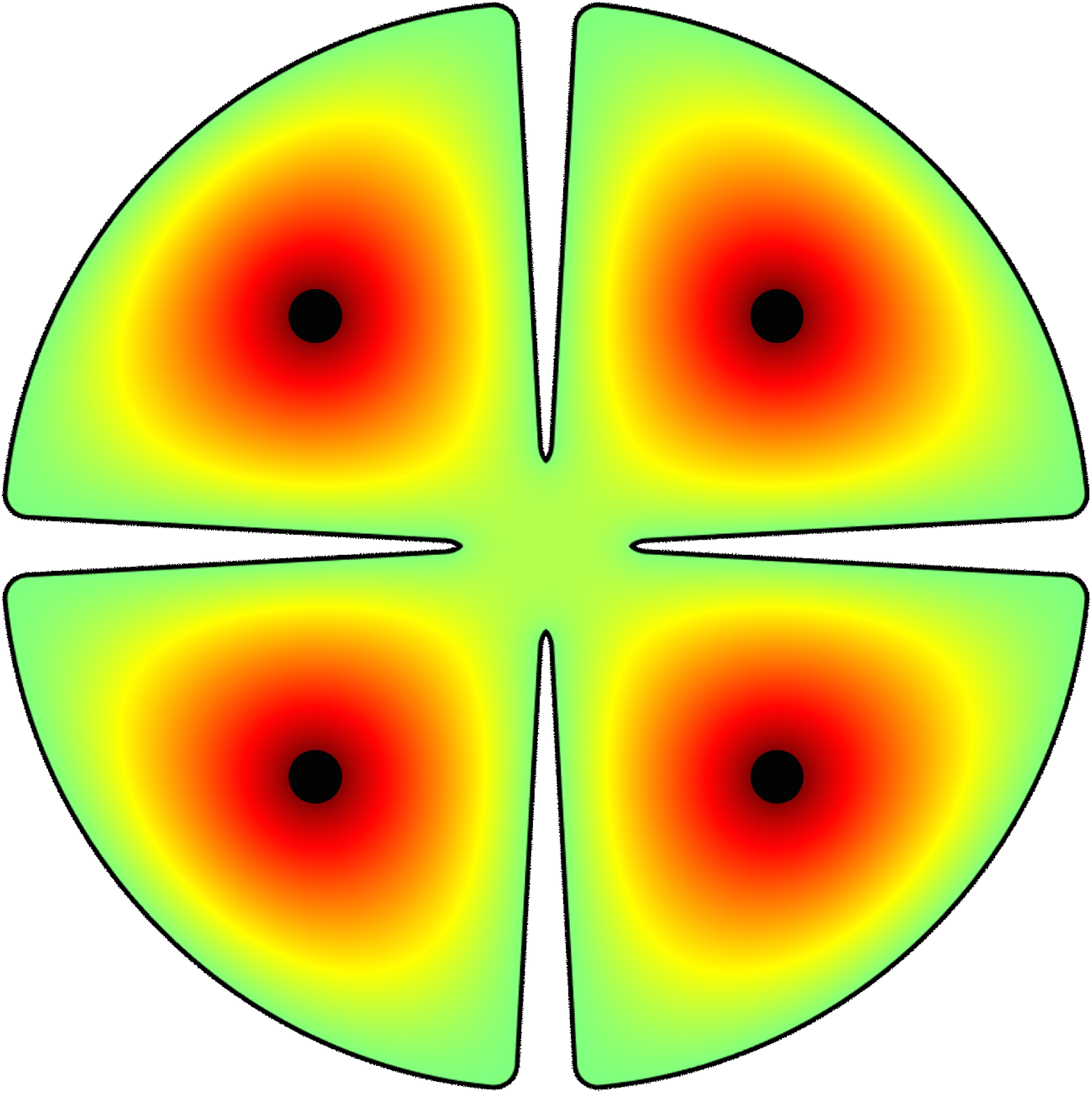}
\hspace{0.01\textwidth}
\includegraphics[width=0.14\textwidth]{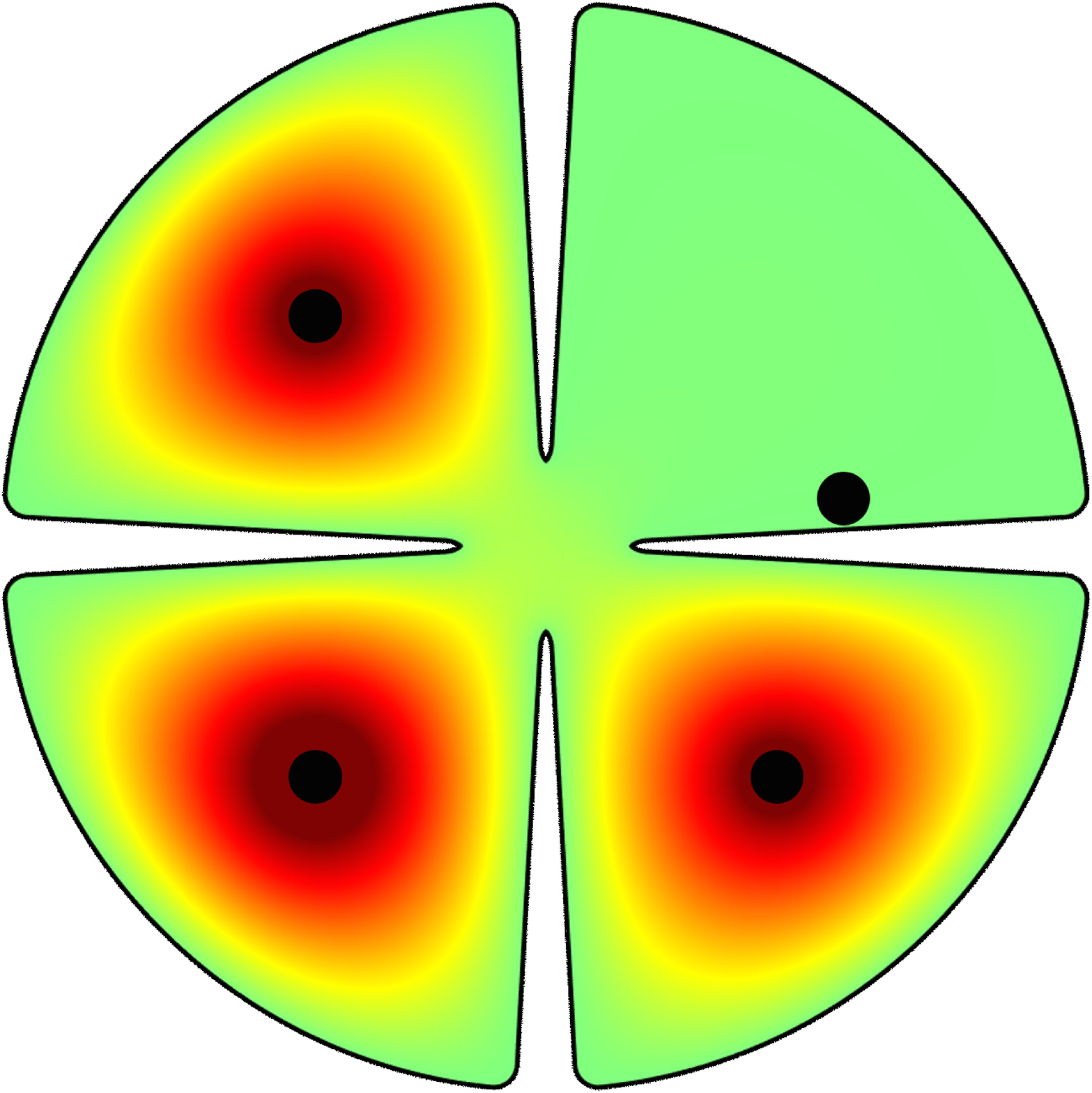}
\hspace{0.01\textwidth}
\includegraphics[width=0.14\textwidth]{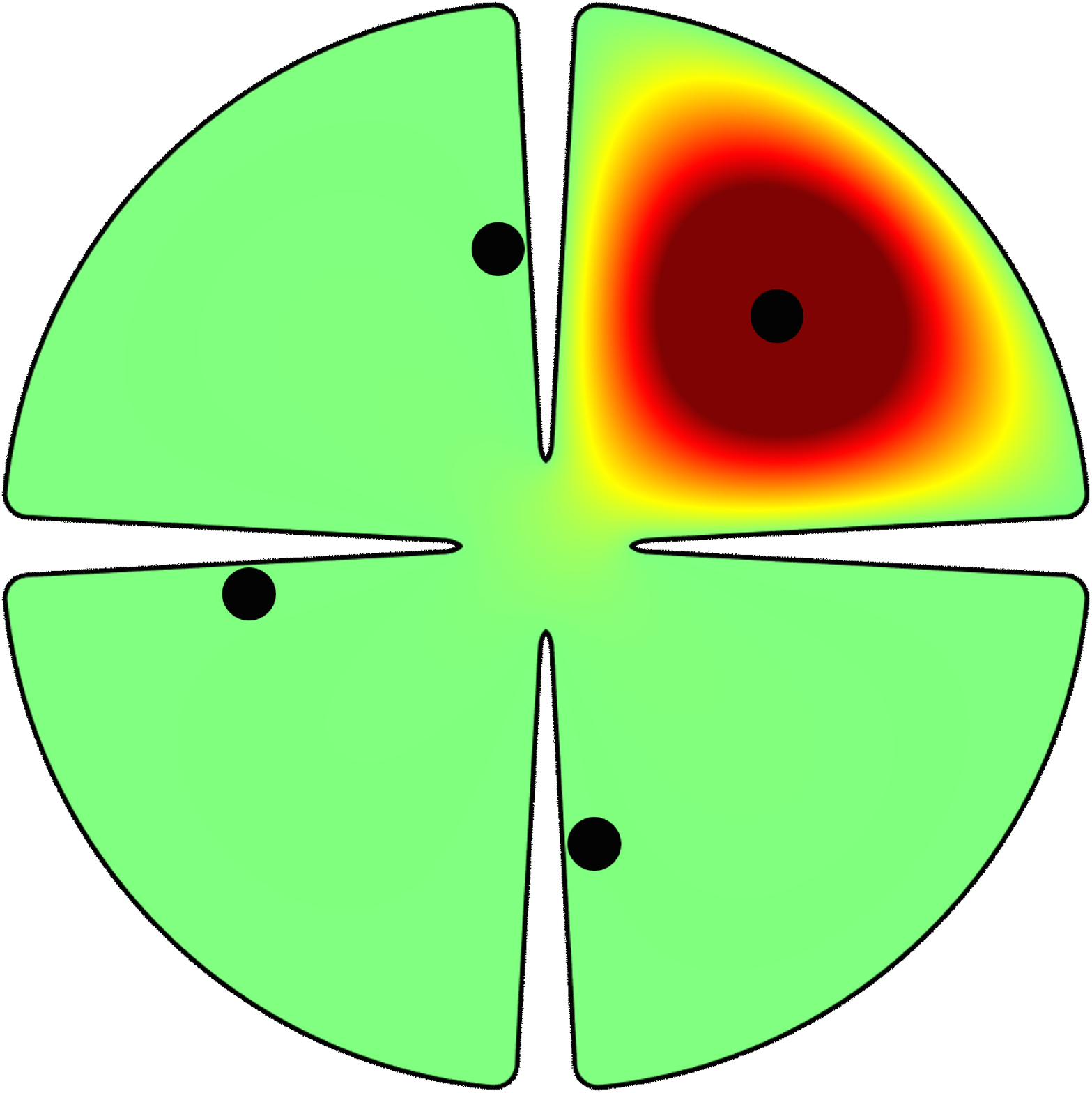}
\caption{Electric field distributions of a quad-cell cavity for different configurations of tuning rod position.
The black bullet in each cell represents the position of the rod.
With the rods are symmetrically aligned, the fields are distributed symmetrically (left), while the field becomes localized with asymmetric rod configurations (middle and right).}
\label{fig:E_field}
\end{figure}

The frequency tuning mechanism for pizza cylinder cavities employs the same concept as used in conventional cylindrical cavity detectors, 
i.e., altering the field distribution of the resonant mode under consideration by moving around a single (pair) of dielectric or metal cylindrical rod(s) inside each cell.
For our simulation study, a lossless dielectric cylindrical tuning rod with a large dielectric constant ($\epsilon_r=10$) is employed.
The size of the tuning rod is optimized from a separate simulation study to be $r/R=0.1$, where $r$ is the rod radius and $R$ is the inner radius of the cylindrical cavity whose TM$_{010}$ resonant frequency is equal to that of the  TM$_{010}$-like mode of the multiple-cell cavity.
Since the individual cells are spatially connected and thus the EM fields can mix, the relative rod position in a cell affects the entire field distribution of the cavity and breaks the field symmetry as illustrated in Fig.~\ref{fig:E_field}.
Therefore, the frequency tuning mechanism for the TM$_{010}$-like mode requires that the field distribution in individual cells be identical and that the overall field distribution be symmetric.
We refer to this condition as ``phase-matching".

\subsection{Electric field profile and coupling}\label{subsec:coupling}
\begin{figure}[h]
\hspace{0.0045\textwidth}
\includegraphics[width=0.155\textwidth]{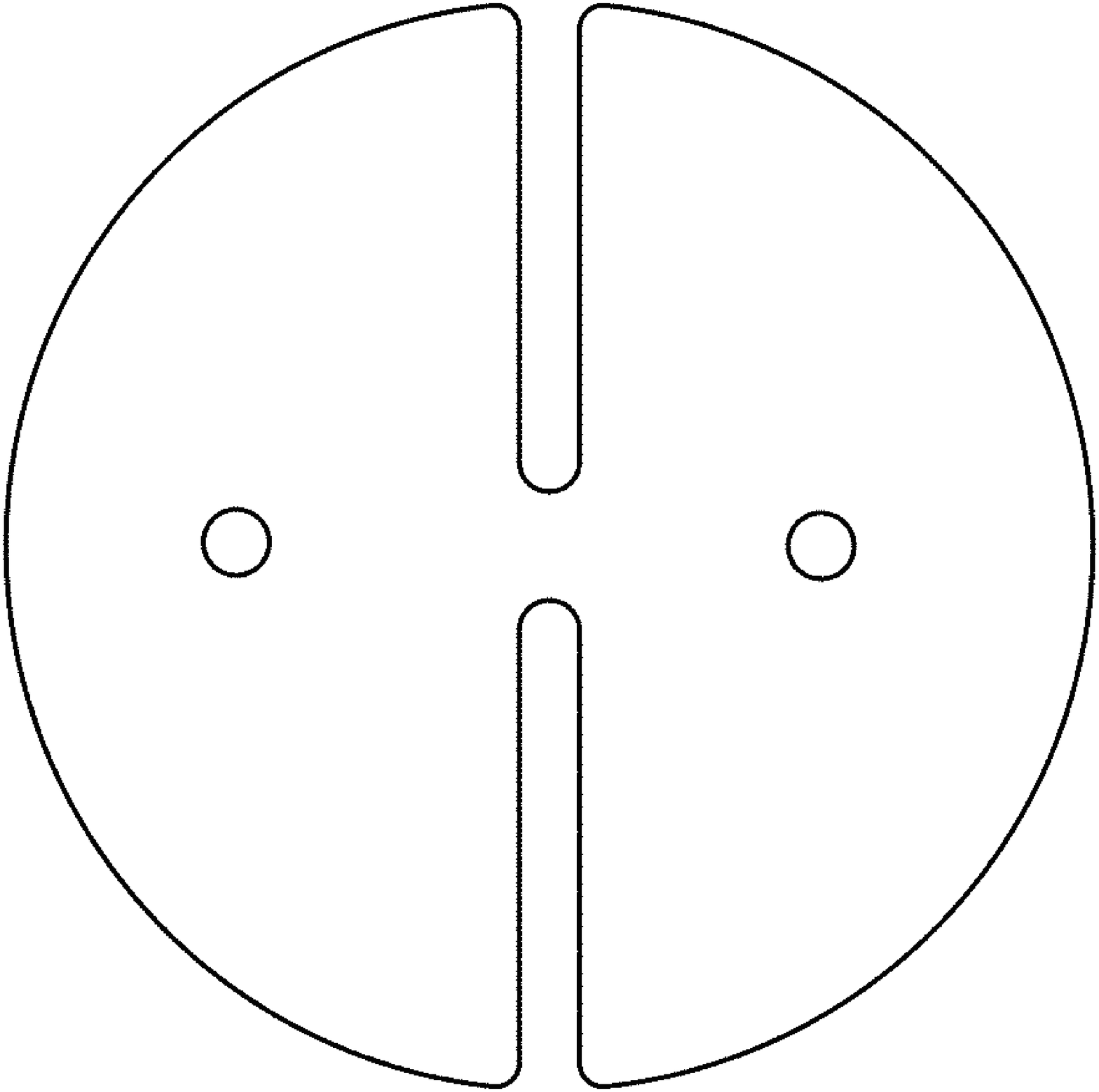}
\includegraphics[width=0.155\textwidth]{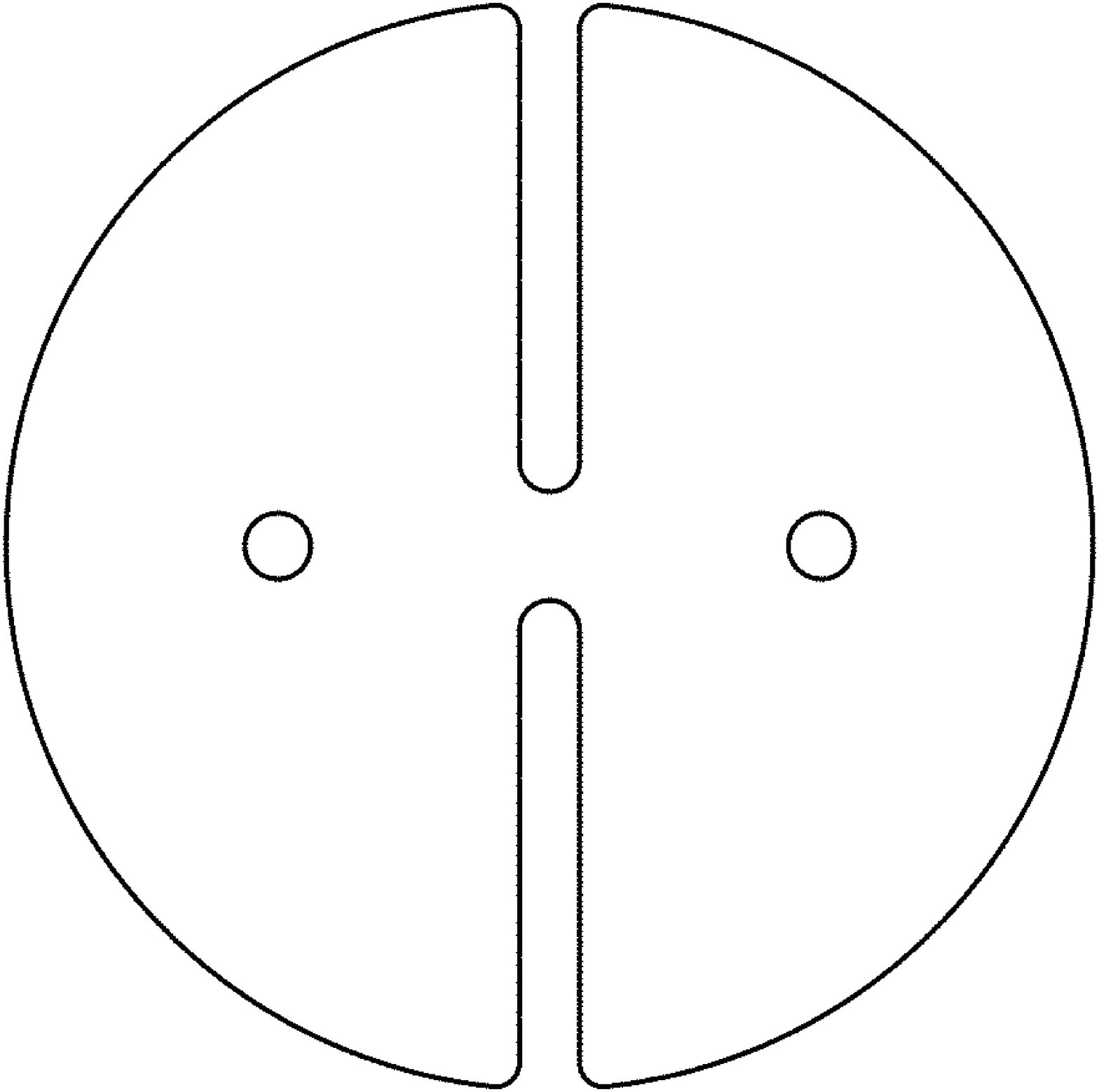}
\includegraphics[width=0.155\textwidth]{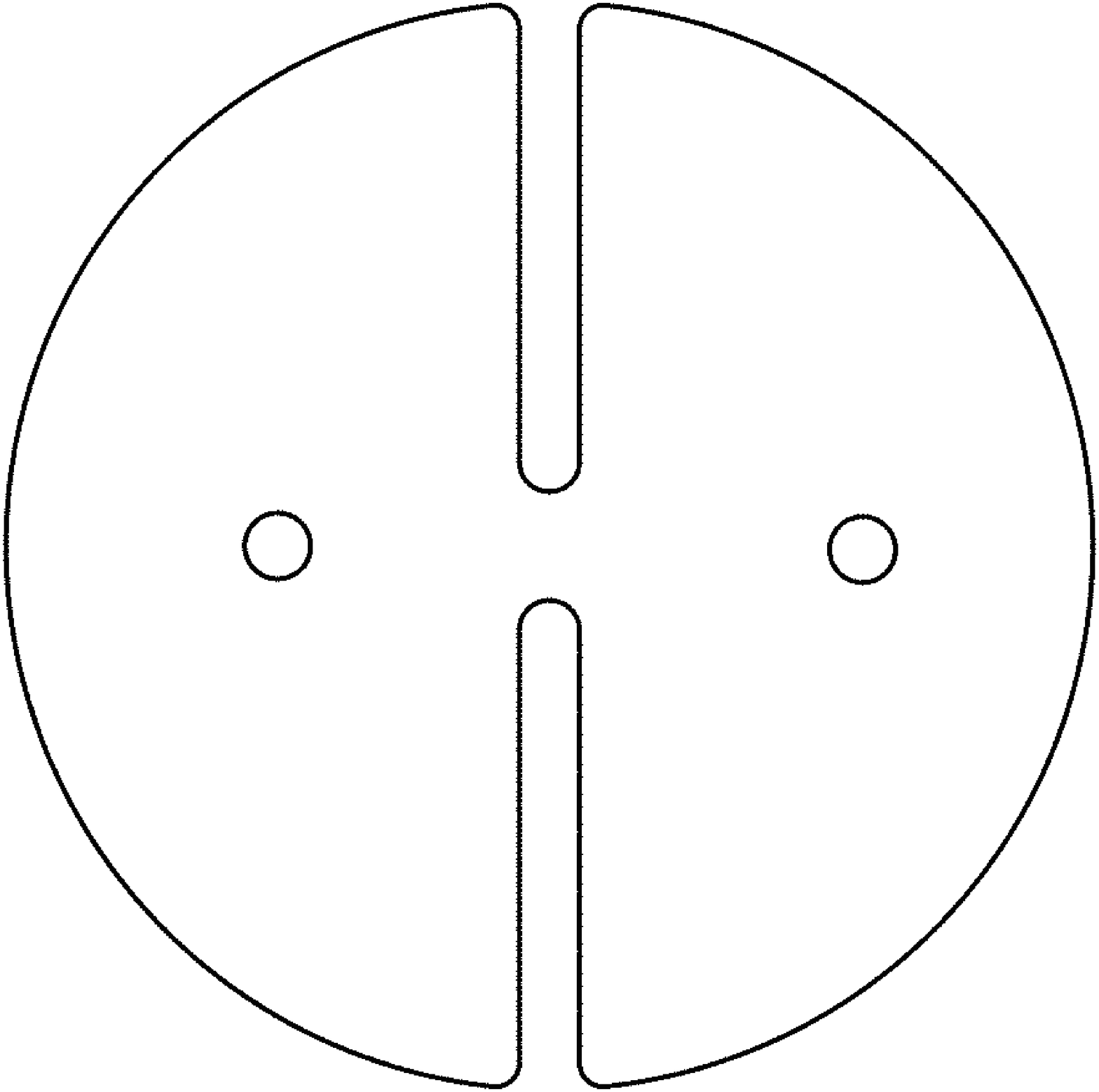}
\subfloat[]{\includegraphics[width=0.16\textwidth]{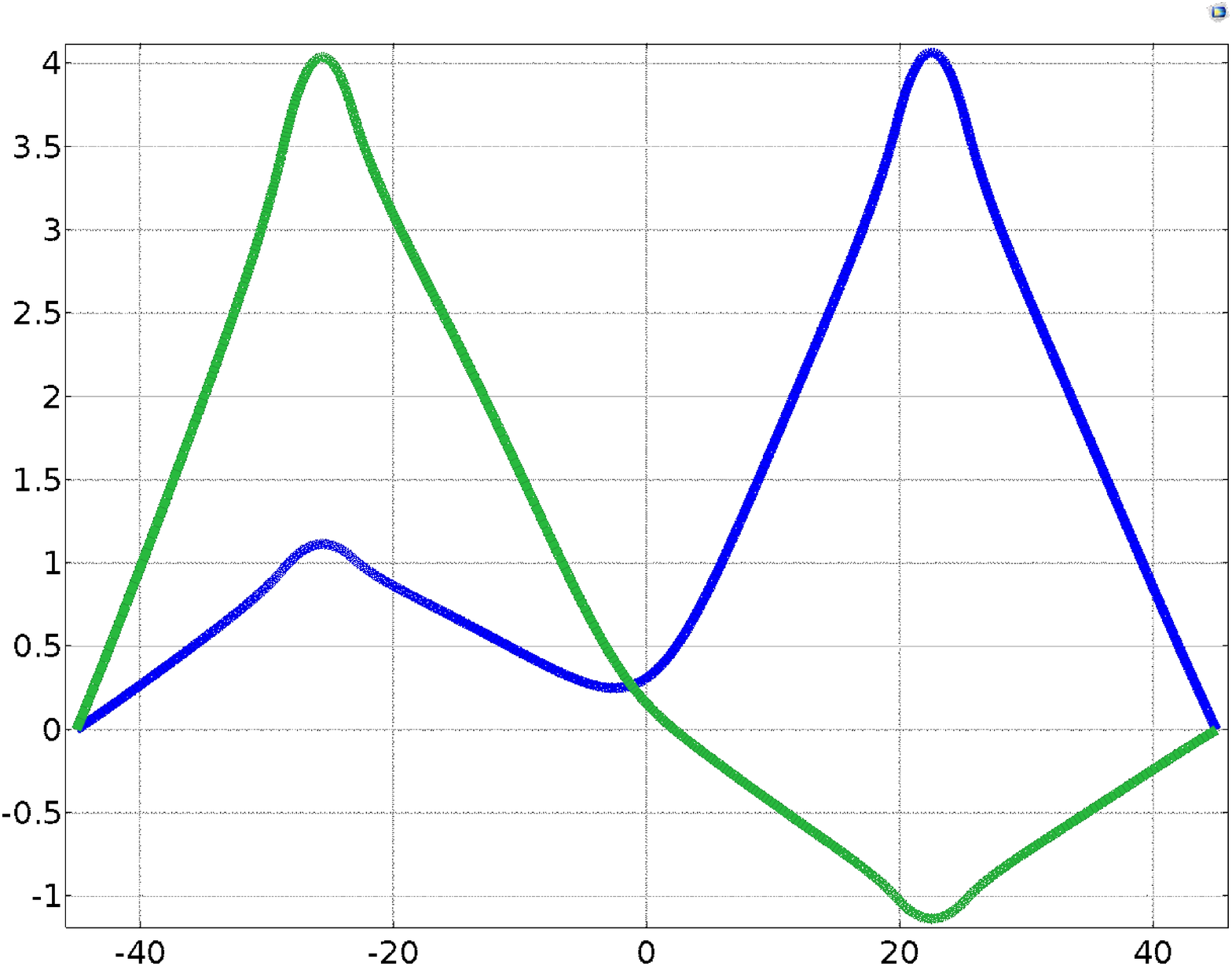}}
\subfloat[]{\includegraphics[width=0.16\textwidth]{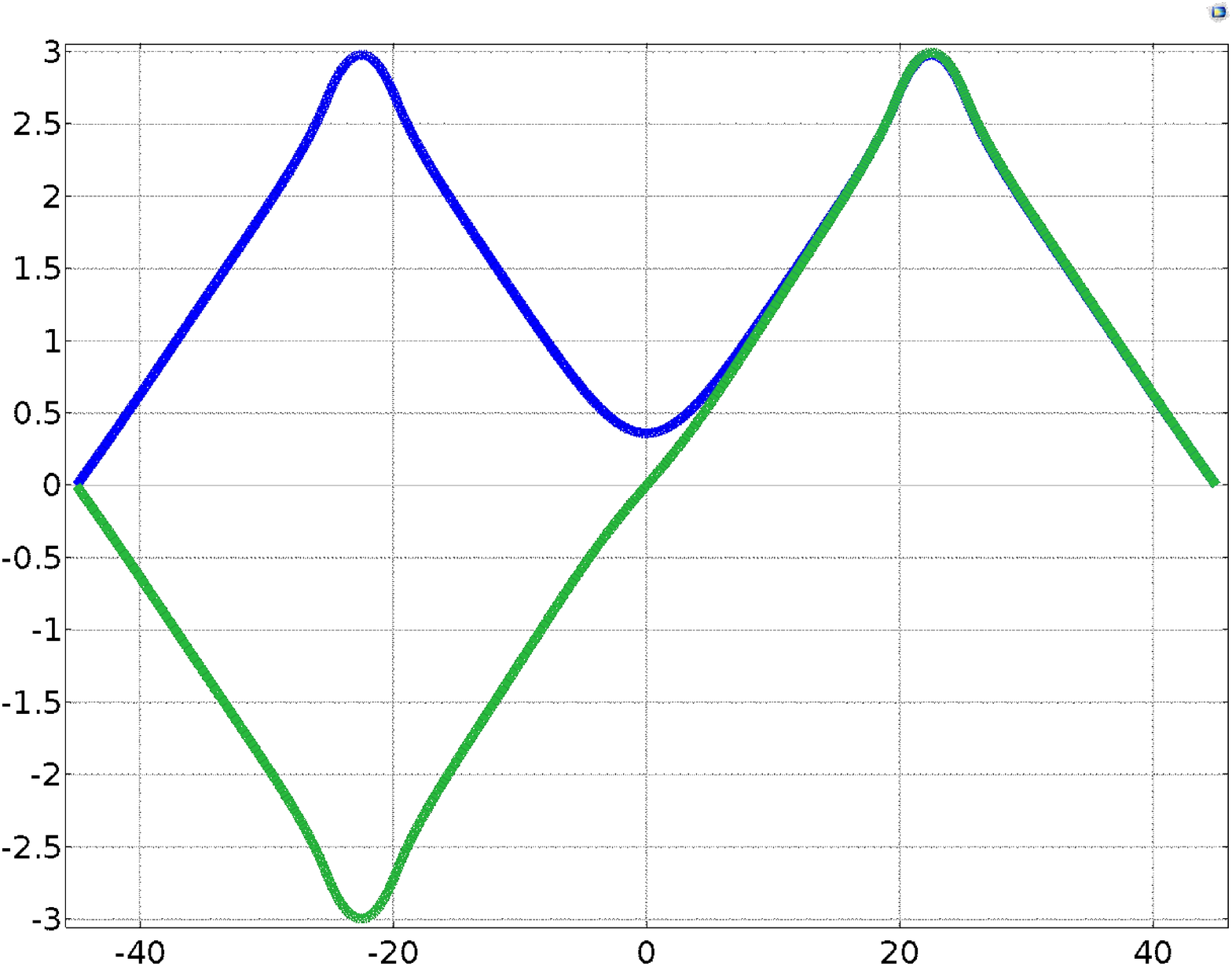}}
\subfloat[]{\includegraphics[width=0.16\textwidth]{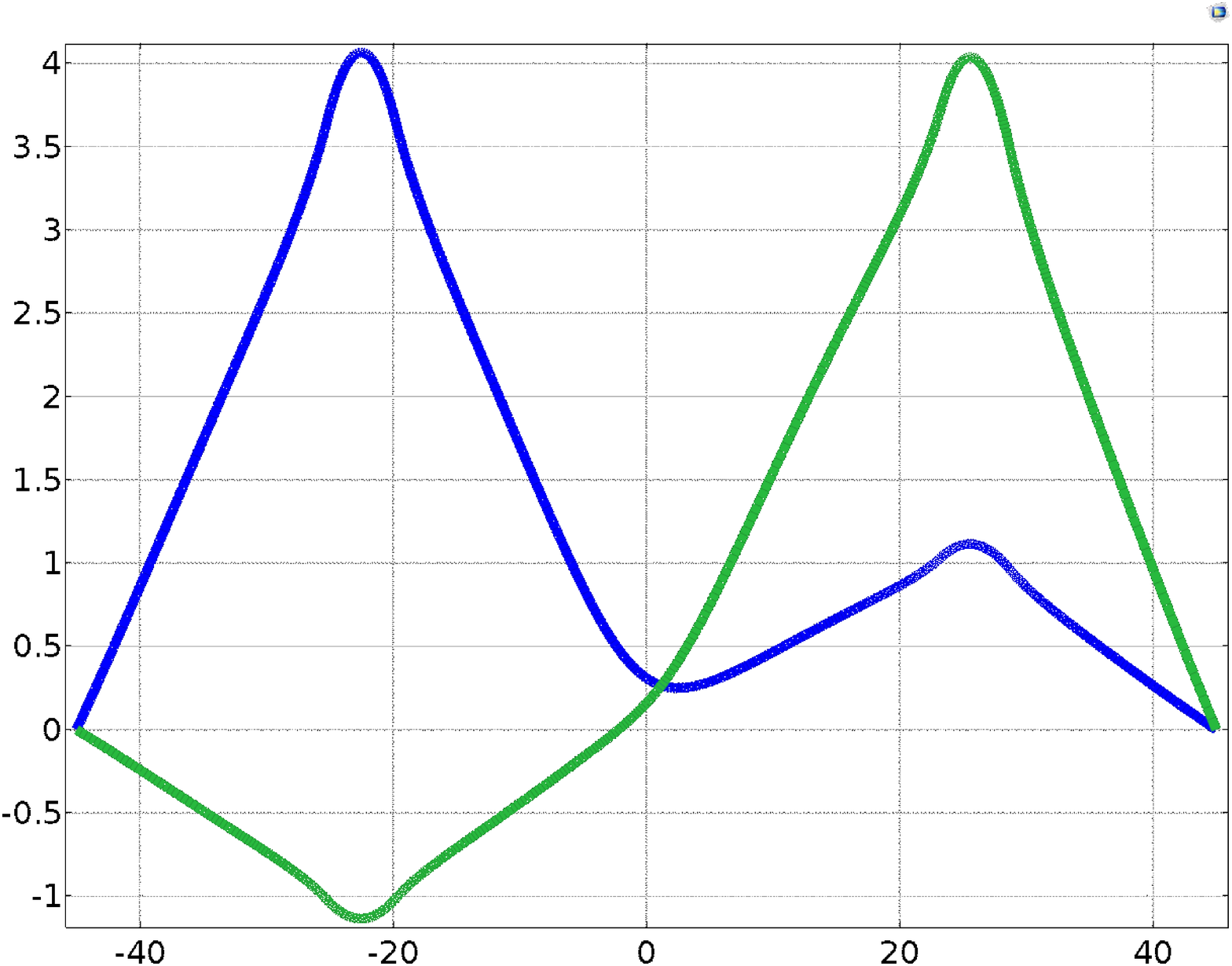}}
\caption{Different rod configurations and corresponding electric field profiles for a double-cell cavity: asymmetric configurations in (a) and (c); and symmetric configuration in (b).
The field profiles of the two resonant modes, the TM$_{010}$-like and TM$_{110}$-like modes, are represented by the blue and green lines respectively.
The difference in rod configurations is more visible in the peak field positions.}
\label{fig:E_profile}
\end{figure}

Now, let us consider the electric field profile in the multiple-cell cavity, which exhibits some interesting features.
For simplicity, we consider a double-cell cavity with a small hollow gap in the middle.
Due to the cell multiplicity, we expect two resonant modes whose frequency degeneracy is broken by the gap: the lowest TM$_{010}$-like mode and the higher TM$_{110}$-like mode.
Figure~\ref{fig:E_profile} shows the electric field profiles of the resonant modes for different configurations of relative rod position.
It can be seen that when the two rods are aligned symmetrically with respect to the center of the cavity, the field distributions of the two modes are also symmetric.
The $E$ fields of the cells are in phase for the TM$_{010}$-like mode, while they are 180 degrees out of phase for the TM$_{110}$-like mode.
This is the condition of phase-matching for multiple-cell cavities.

It is noticeable that once phase-matching is successfully accomplished, as in Fig.~\ref{fig:E_profile}(b), 
the field strength at the center of the cavity becomes zero for the TM$_{110}$-like mode, while it remains non-zero for the TM$_{010}$-like mode.
This indicates that electrical coupling at the center of the cavity will be sensitive only to the lowest TM$_{010}$-like mode, not to any higher modes\protect\footnotemark.
This, in turn, provides an idea that a single monopole RF antenna in the middle of the cavity is sufficient to extract the signal from the entire detection volume.
It is also found that vane-shape partitions, as shown in Fig.~\ref{fig:E_field}, enhance the field strength at the center of the cavity, because the fields are pushed toward it,
while the field strength remains zero for the higher modes.
\footnotetext{This feature is common for multiple-cell cavities with even multiplicity.}

\subsection{Determination of the gap size}
Based on observations from the simulation studies, the hollow size is optimized by imposing two criteria:
\begin{itemize}
\item $\Delta f / \delta f > 5$, where $\Delta f$ is the frequency difference between the two lowest modes and $\delta f$ is the full width at half maximum of the Lorentzian distribution of the cavity response (e.g., scattering parameter) for the lowest TM$_{010}$-like mode; and
\item $E_{\rm{cen}}/E_{\rm{max}}>0.1$, where the ratio is the relative $E$ field strength at the center of the cavity  ($r$=0)  to the maximum field strength for the TM$_{010}$-like mode.
\end{itemize}
The first criterion is for the mode separation, while the second one is related to the signal coupling.
From a simulation study, we obtain an optimal hollow size for quadruple-cell cavities of typical copper of $d/D=0.1$, where $d$ is the gap size and $D$ is the cavity inner diameter.
The power degradation due to the gap is found to be only about 2\%.

\subsection{Scattering parameter and Smith chart}
\begin{figure}[h]
\centering
\includegraphics[width=0.155\textwidth]{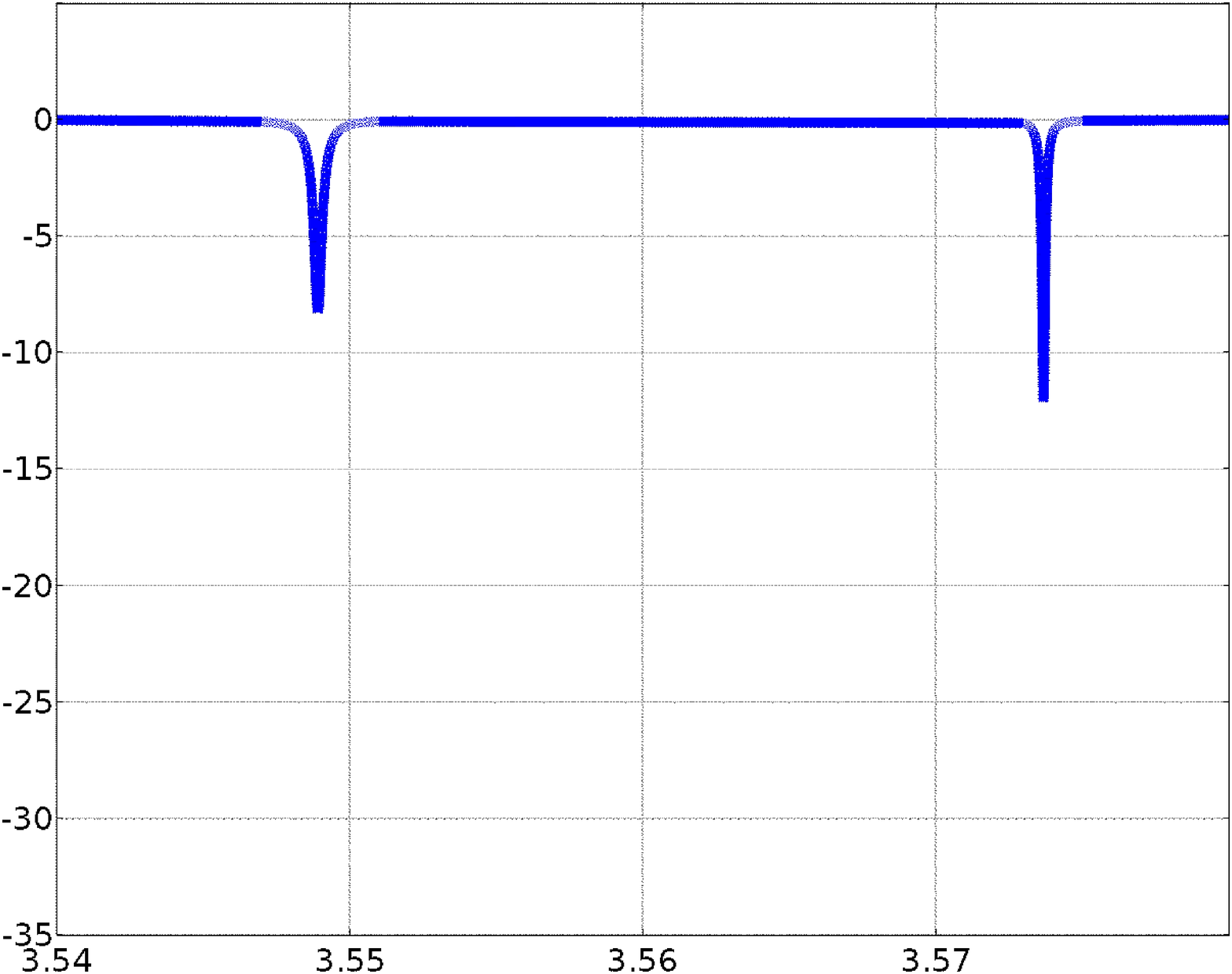}
\includegraphics[width=0.155\textwidth]{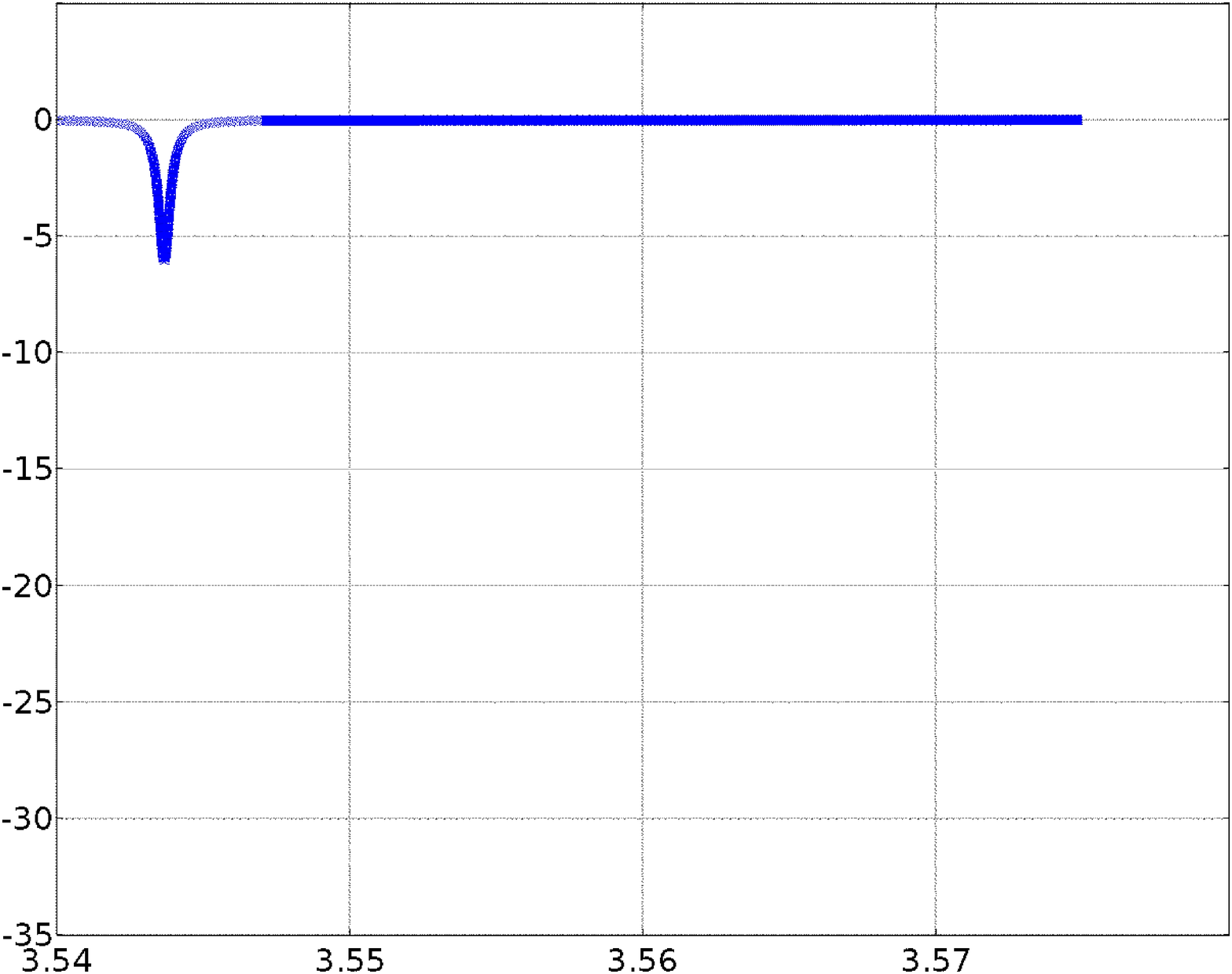}
\includegraphics[width=0.155\textwidth]{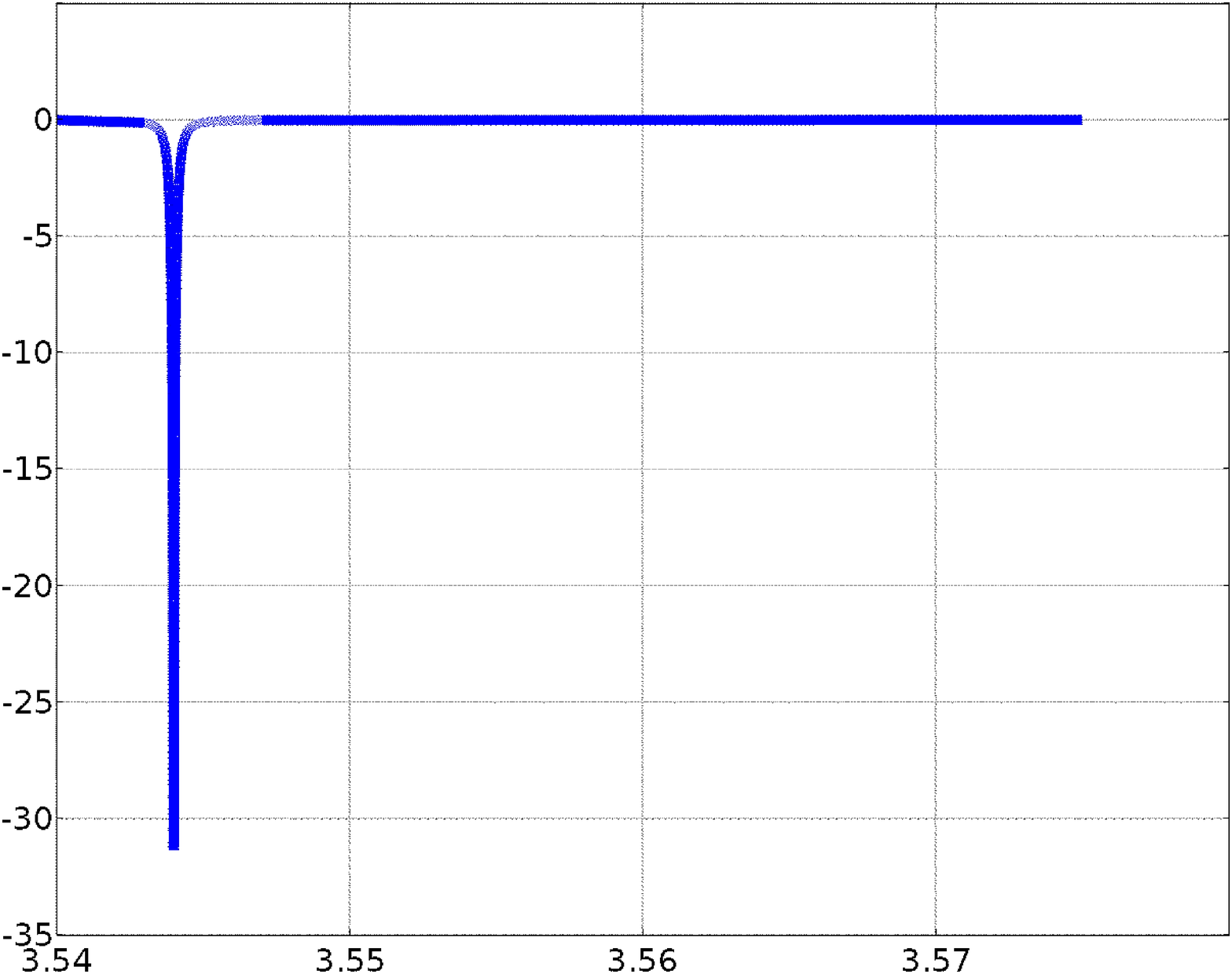}
\subfloat[]{\includegraphics[width=0.1625\textwidth]{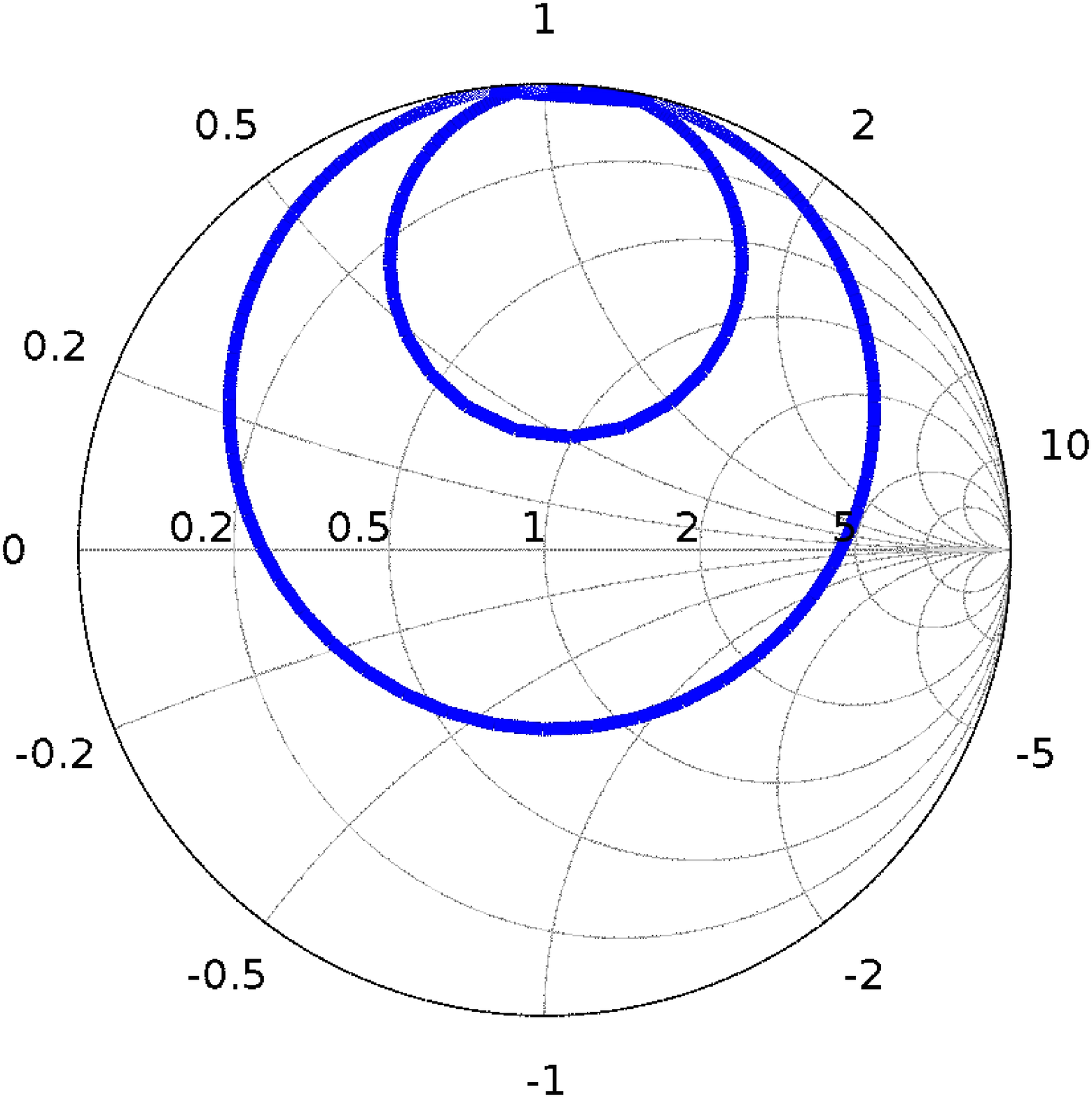}}
\subfloat[]{\includegraphics[width=0.1625\textwidth]{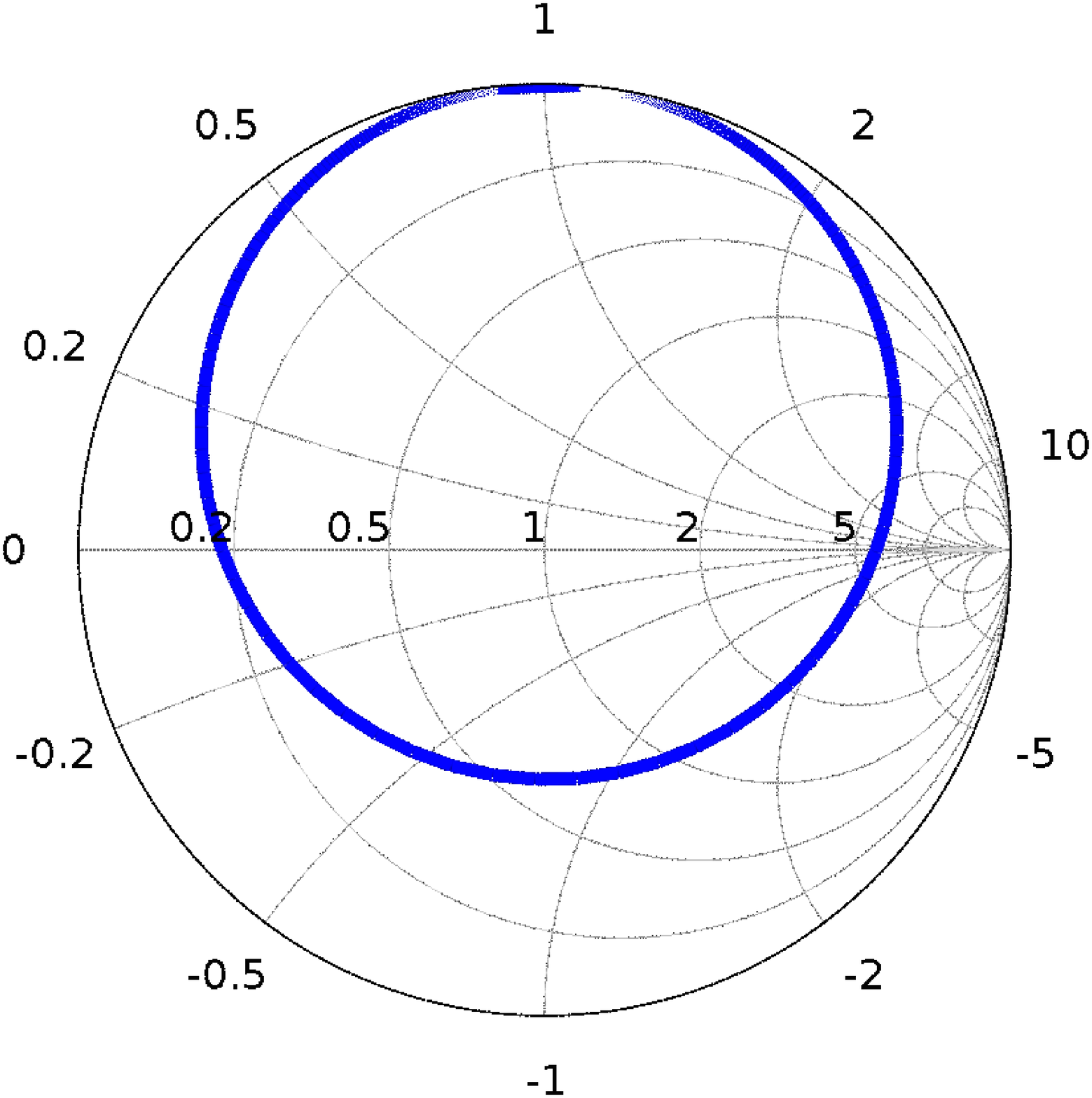}}
\subfloat[]{\includegraphics[width=0.1625\textwidth]{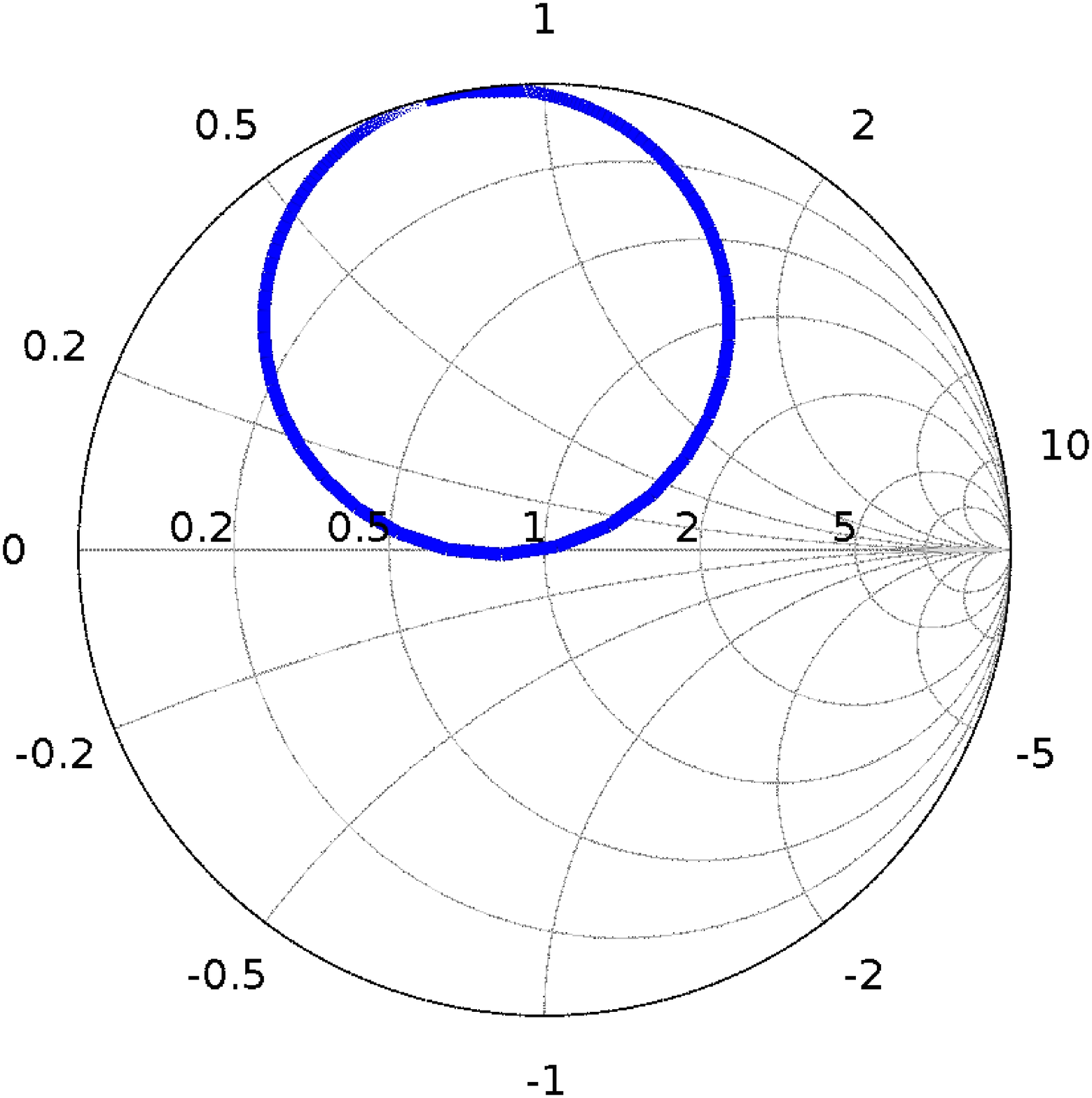}}
\caption{Representation of phase-matching and critical coupling in the $S$-parameter and Smith chart for a double-cell cavity: 
(a) arbitrary rod configuration and coupling; 
(b) symmetric rod configuration (phase-matching) and arbitrary coupling; 
and (c) symmetric rod configuration and critical coupling.}
\label{fig:S11_Smith}
\end{figure}

The features described in Sec.~\ref{subsec:degeneracy}$\sim$\ref{subsec:coupling} are in practice represented in the scattering parameter ($S$-parameter) and the Smith chart in a network analyzer.
Figure~\ref{fig:S11_Smith} shows examples of the frequency spectra of the $S$-parameter associated with reflection and the corresponding constant resistance circles in the Smith chart for the different rod configurations and coupling strengths for a double-cell cavity.
An asymmetric rod configuration and arbitrary coupling strength are represented by two reflection peaks in the $S$-parameter space and two circles in the Smith chart.
When the rods are symmetrically arranged, the higher resonant frequency peak and the corresponding circle fade away.
This implies that phase-matching of multiple-cell cavities is characterized by disappearance of any higher mode peaks in the scattering parameter space and corresponding constant resistance circles in the Smith chart.
Once phase-matching is successfully achieved, an RF antenna on the top (or bottom) of the cavity couples only to the non-zero field lowest mode at the center.
The depth of insertion of the antenna into the cavity is adjusted until the critical coupling condition, defined as the reflection coefficient $\Gamma = \frac{|1-\beta|}{|1+\beta|}=0$ for the coupling strength $\beta$, is fulfilled.
In this condition, the remaining constant resistance circle passes through the center of the Smith chart.

\section{Tuning mechanism}
\begin{figure}[h]
\centering
\includegraphics[width=0.35\textwidth]{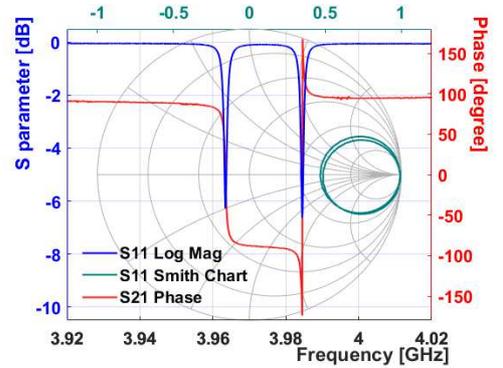}
\caption{(Color online.) Frequency spectra of various $S$-parameter measurements using a vector network analyzer under the condition of phase-matching described in the text already implemented. 
The blue line measures the reflection through an antenna arbitrarily coupled to a cell - two peaks correspond to the TM$_{010}$-like and TM$_{020}$-like modes.
The green circles are the representation of the two reflection peaks in the Smith Chart.
The compatible magnitude implies that the stored energy in the cell is equally shared by the two modes.
The red line shows the phase information of a transmitted signal between the two cells using a pair of identical antennae.
After calibration using a reflected signal, the absolute phase reads $-7.8^\circ$ and $-174.1^\circ$ at the two resonant frequencies, indicating that the two cells resonate in phase for the lower mode and out of phase for the higher mode.
}
\label{fig:phase_matching}
\end{figure}

The tuning system of a multiple($N$)-cell cavity detector is comprised of $N$ identical dielectric rods for individual cells and a single coupler in the middle of the cavity.
The basic principle of the tuning mechanism is the same as that for conventional multiple-cavity systems, and consists of two steps: phase-matching followed by critical coupling.
From an experimental point of view, phase-matching is achieved by aligning the tuning rods until the higher mode peaks in the $S$-parameter space and, equivalently, the corresponding constant resistance circles in the Smith chart vanish.
For our application, we require the coupling strength to the higher modes to be $S_{11}<0.05$\,dB at the sacrifice of conversion power loss less than 1\%.
This condition of the phase-matching is validated by examining the $S$-parameters using a vector network analyzer as depicted in Fig.~\ref{fig:phase_matching}.

The coupling mechanism is simple enough that a single coaxial monopole antenna is placed at the center of the top (or bottom) end cap and its depth into the cavity changes the coupling strength to the lowest TM$_{010}$-like mode independently of the other higher modes.
Critical coupling is characterized by the $S$-parameter associated with reflection, $S_{11}=-10\log\Gamma^2$ becoming negative infinity or practically $S_{11} <-30$\,dB, and equivalently by the constant resistance circles passing through the center of the Smith chart.
Using the 3 dB method, the resonant frequency and the cavity $Q_L$ are determined from the reflection $S$-parameter of the lowest mode peak.
Figure~\ref{fig:S11_Smith} shows the sequence of the tuning mechanism; features in each step are represented by reflection peaks in the scattering parameter space and constant resistance circles on the Smith chart.

\section{Demonstration of experimental feasibility}
\begin{figure}[h]
\centering
\subfloat[]{\includegraphics[width=0.11\textwidth]{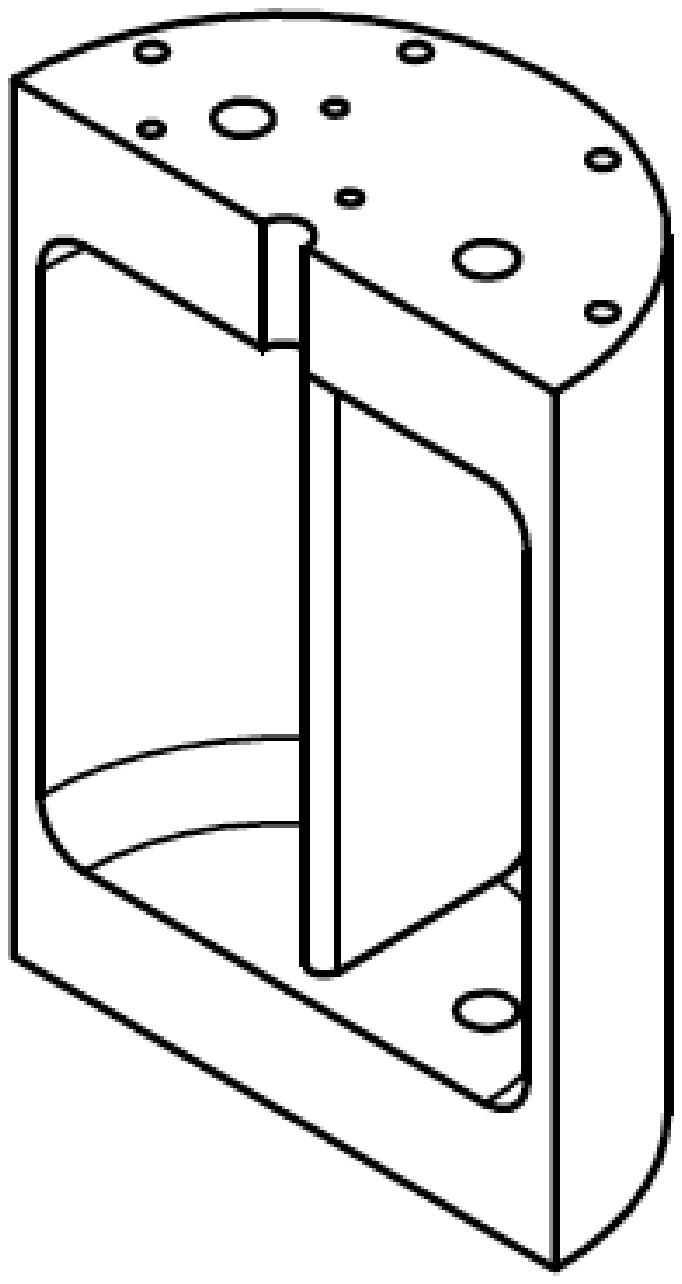}}
\hspace{0.05\textwidth}
\subfloat[]{\includegraphics[width=0.25\textwidth]{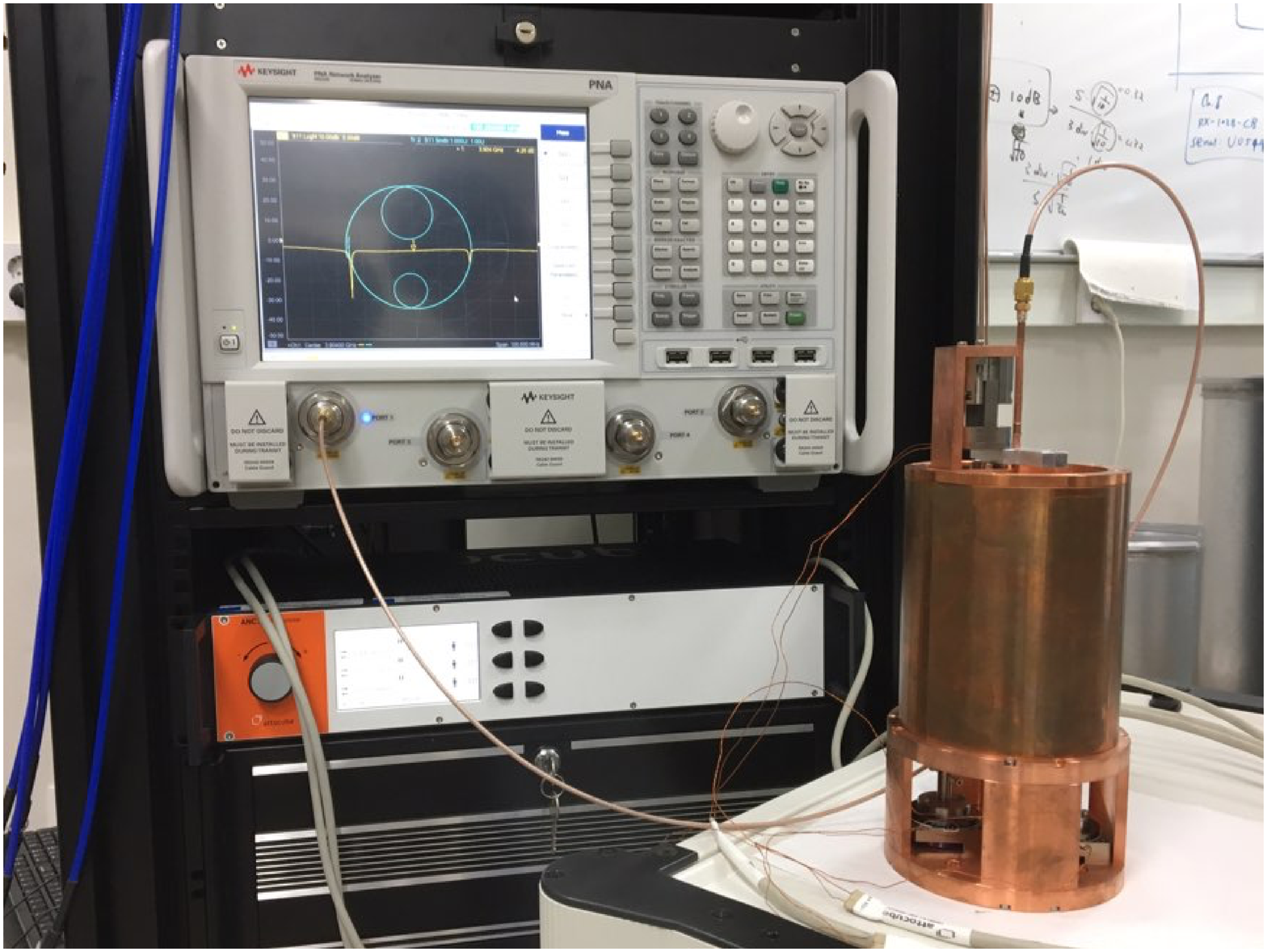}}
\caption{(a) Drawing of half piece of split cavity. The partition in the middle divides the cavity into two cells when it is assembled.
(b) Photo of the system setup consisting of a copper cavity, two rotators under the cavity, a single antenna above the cavity, an actuator controller and a network analyzer.}
\label{fig:split}
\end{figure}

\begin{figure}[h]
\centering
\includegraphics[width=0.35\textwidth]{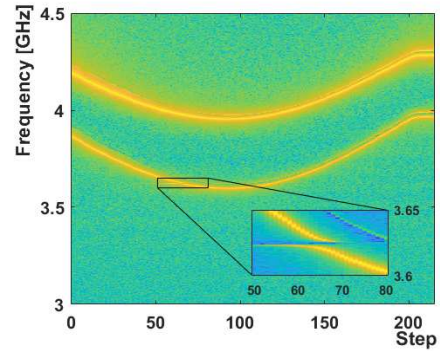}
\caption{(Color online.) Mode map of the double-cell cavity described in the text. 
The lower and higher modes correspond to the TM$_{010}$-like and TM$_{011}$-like modes, respectively.
The frequency tuning range for the TM$_{010}$-like mode is about 400\,MHz ($3.6\,\rm{GHz}\sim4.0\,GHz$).
A mode mixing with a TE mode, identified as the TE$_{211}$-like mode from simulation, is seen at around 3.625 GHz in the zoomed plot.}
\label{fig:modemap}
\end{figure}

In order to verify the experimental feasibility of the multiple-cell design, a demonstration is performed using a double-cell cavity at room temperature.
The cavity is made of oxygen-free high conductivity copper (OFHC) with 99.99\% nominal purity and dimensions of 90\,mm inner diameter, 100\,mm inner length, 5\,mm wall thickness, and 10\,mm hollow gap.
The resonant frequency of the TM$_{010}$-like mode is 4.2 GHz and its unloaded quality factor is 13,500.
The concept of split cavity design, introduced in Ref.~\cite{bib:magnetoresistance}, is adopted to eliminate the contact resistance, which is a common problem in for conventional cylindrical cavities.
The cavity consists of two identical pieces of vertically cut half-cylinder, which are hollowed out except for a partition in the middle, as can be seen in Fig.~\ref{fig:split} (a), tied up with copper flanges.
A single tuning rod, made of 99.5\% aluminium oxide (Al$_2$O$_3$), is introduced in each cell and is translated by a rotational piezoelectric actuator installed below the cavity.
A single coaxial RF antenna is inserted through a hole at the top center of the cavity and the depth into the cavity is adjusted by a linear piezoelectric actuator installed above the cavity.
The overall system setup is shown in Fig.~\ref{fig:split} (b) and the mode map of the system is found in Fig.~\ref{fig:modemap}.

\begin{figure}[h]
\centering
\includegraphics[width=0.238\textwidth]{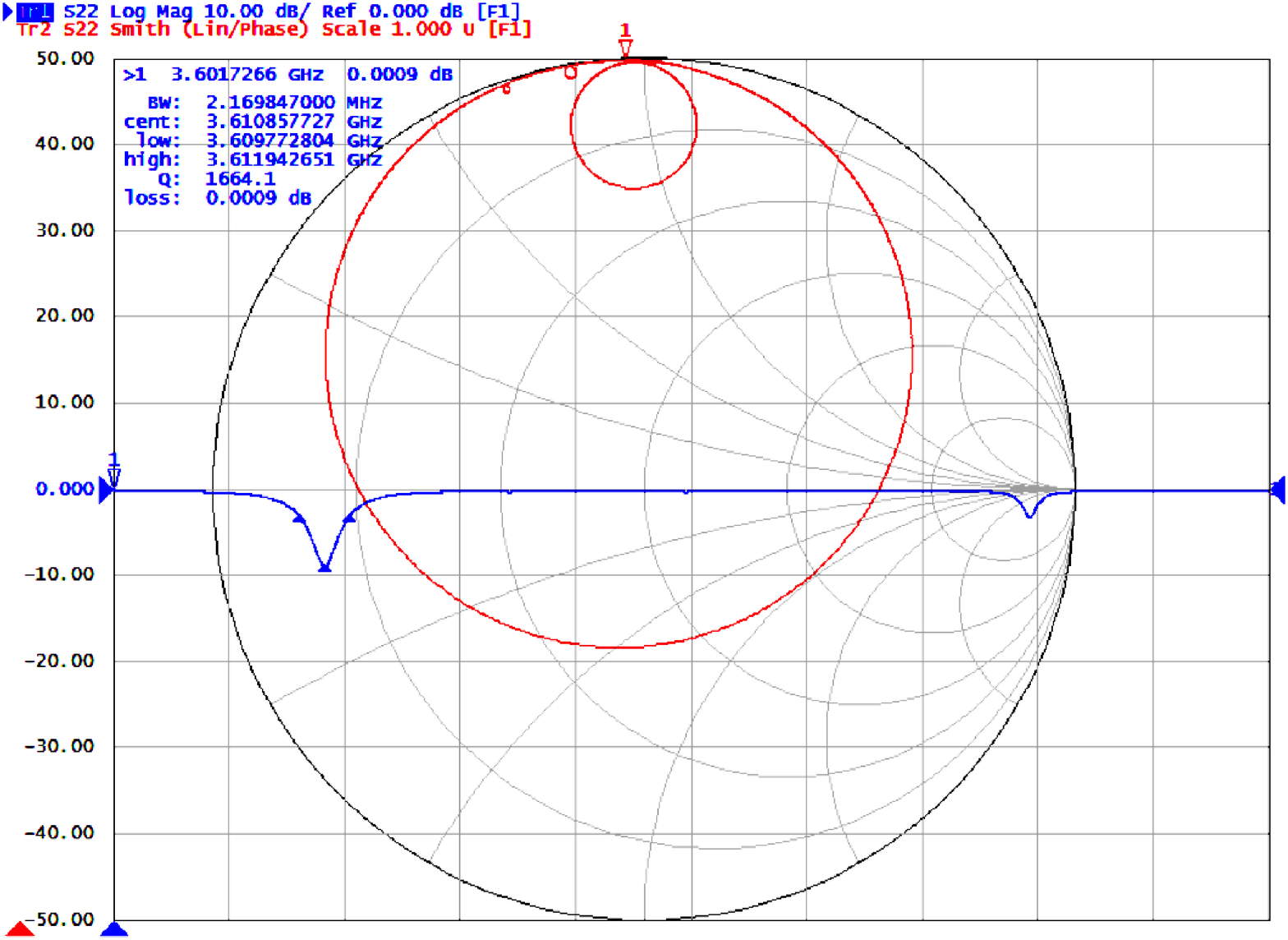}
\includegraphics[width=0.238\textwidth]{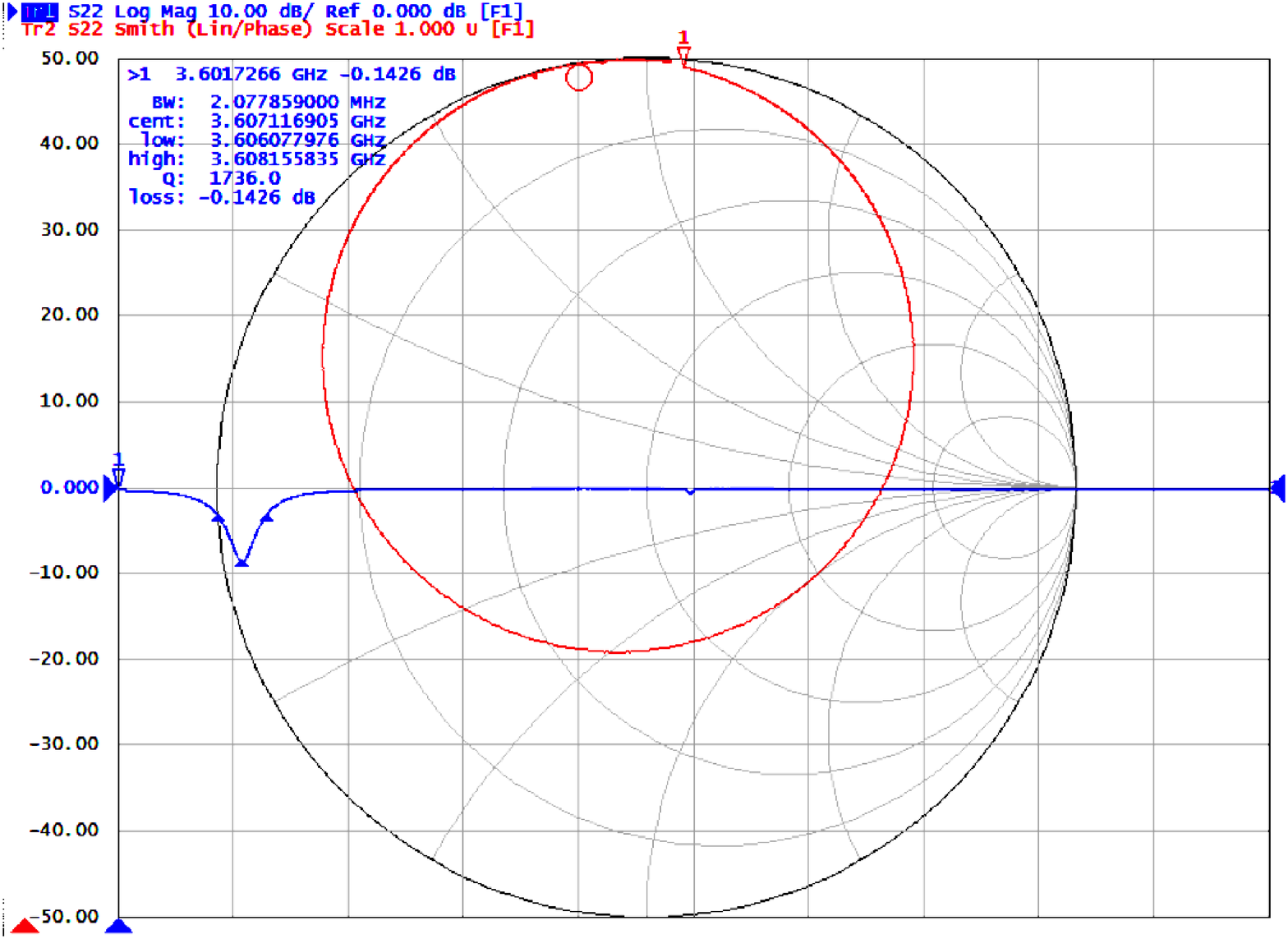}
\includegraphics[width=0.238\textwidth]{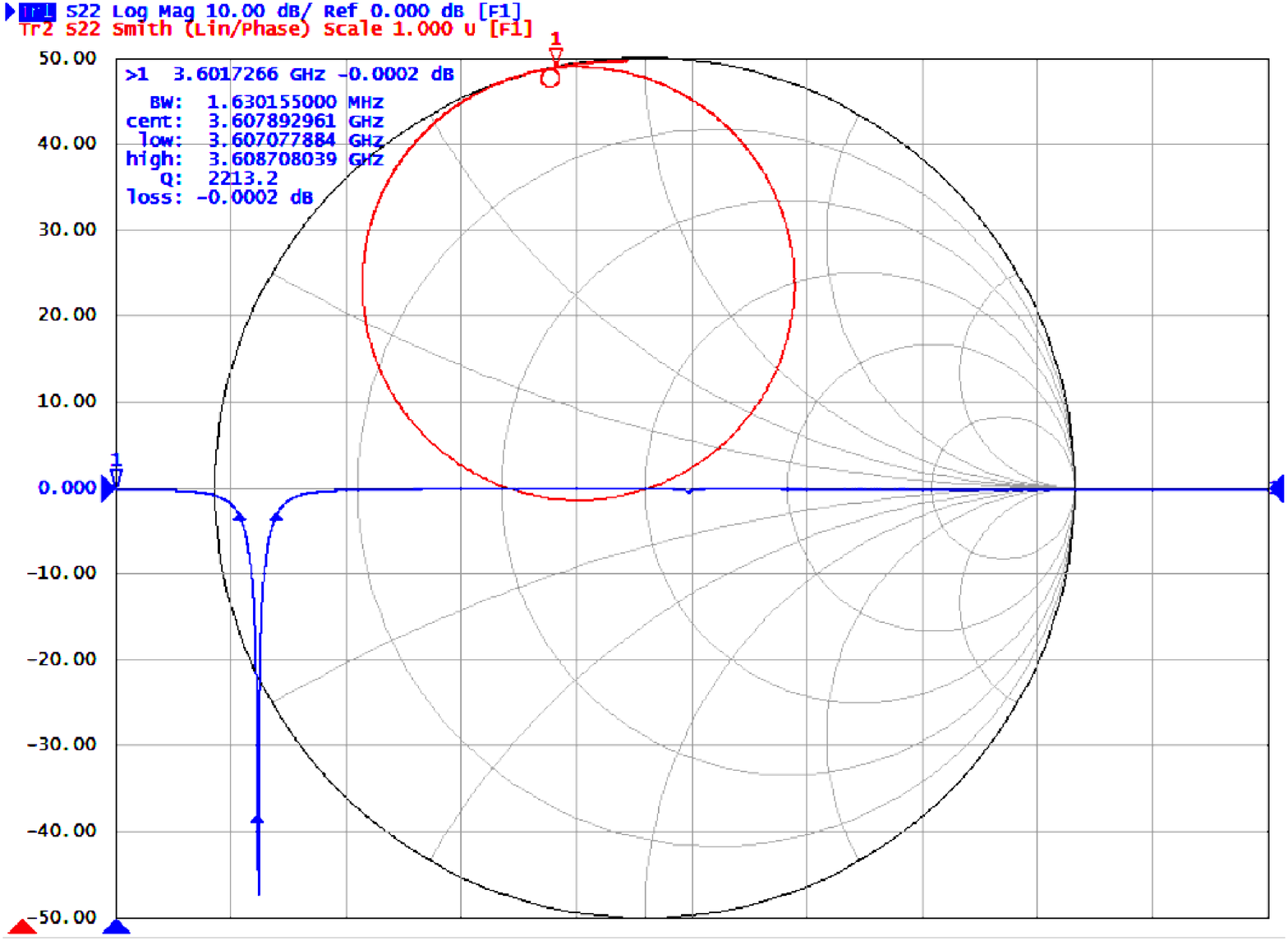}
\caption{(Color online.) Experimental demonstration of the tuning mechanism for a double-cell cavity.
Blue lines are the scattering parameter spectra associated with reflection in logarithmic scale and red circles are their representation in the Smith chart.
Two resonant modes are featured by the two reflection peaks in the $S$-parameter spectrum and the two circles in the Smith chart (top left)\protect\footnotemark.
Phase-matching is assured by vanishing of the higher frequency peak and the corresponding circle (top right).
Critical coupling is characterized by the reflection peak of $S_{11}<-40$\,dB and the remaining circle passing through the center of the chart (bottom).}
\label{fig:demo}
\end{figure}

In the first stage, the tuning rods are placed at slightly different positions in the individual cells and the RF antenna is arbitrarily coupled with the system.
Due to the hollow gap in the middle of the cavity, the frequency degeneracy is seen to be already broken.
This initial state is represented by two peaks in the scattering parameter space, where the lowest and higher frequency peaks correspond to the TM$_{010}$-like and TM$_{110}$-like modes respectively, and by two circles in the Smith chart none of which passes through the center of the chart.
For phase-matching, one of the rotational piezoelectric actuators is manipulated to finely move the tuning rod until the higher frequency peak vanishes.
We impose the criteria that the amplitude of the peak must be less than 0.05\,dB.
It can be noticed that one of the circles in the Smith chart becomes unnoticeably small.
At this point, we assume that the system is phase-matched and that only the TM$_{010}$-like mode is coupled to the antenna.
Finally, the linear piezoelectric actuator is operated to adjust the strength of the coupling until the constant resistance circle passes through the center of the Smith chart and, equivalently, 
 the reflection peak reaches a value below $-40$\,dB in the $S$-parameter space.
The sequence of the demonstration and the characteristics of each step are illustrated in Fig.~\ref{fig:demo}.

\footnotetext{Additional tiny peaks and circles in-between are due to TE modes which are ignored in this study.}

\begin{figure}[h]
\centering
\includegraphics[width=0.35\textwidth]{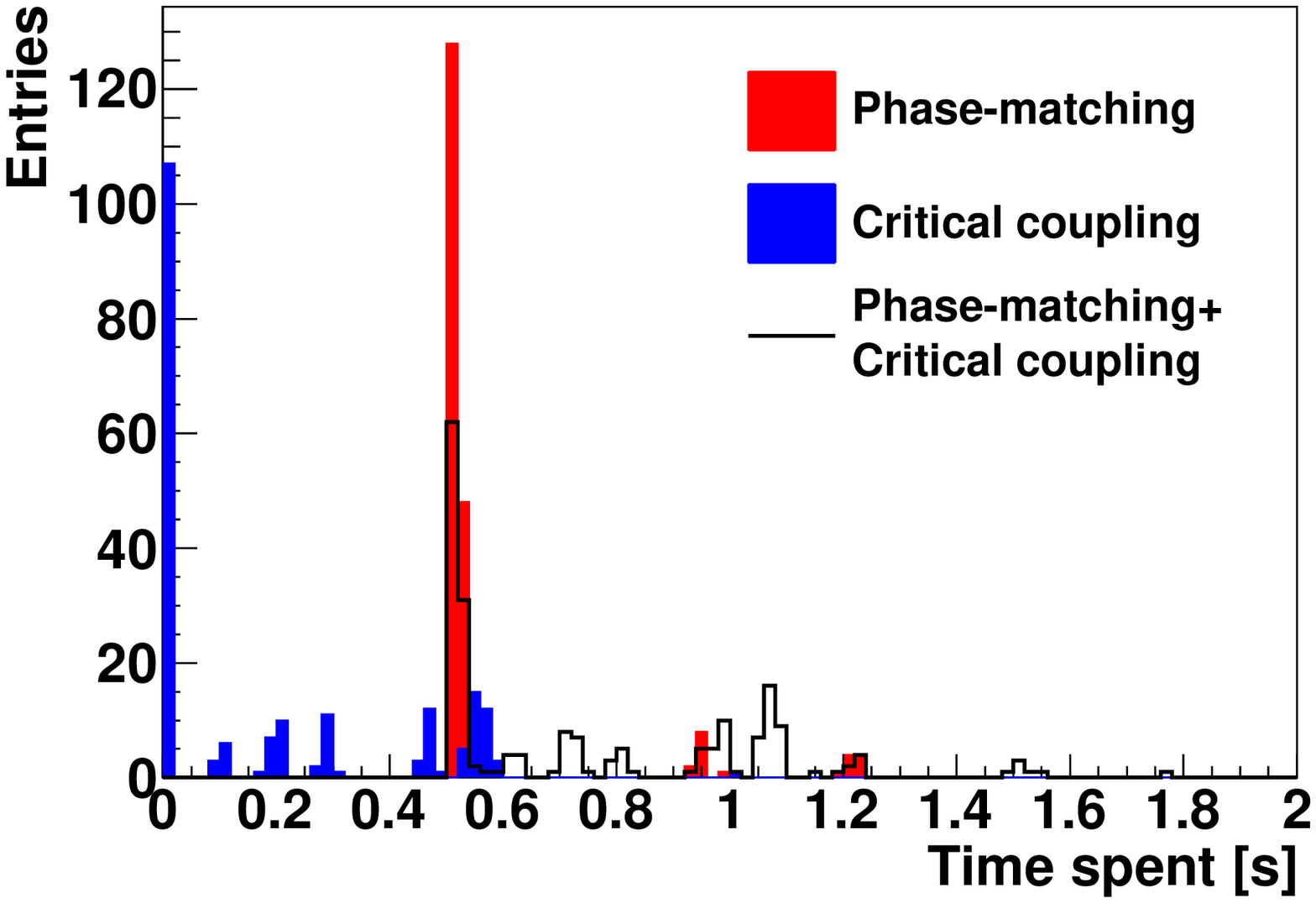}
\includegraphics[width=0.35\textwidth]{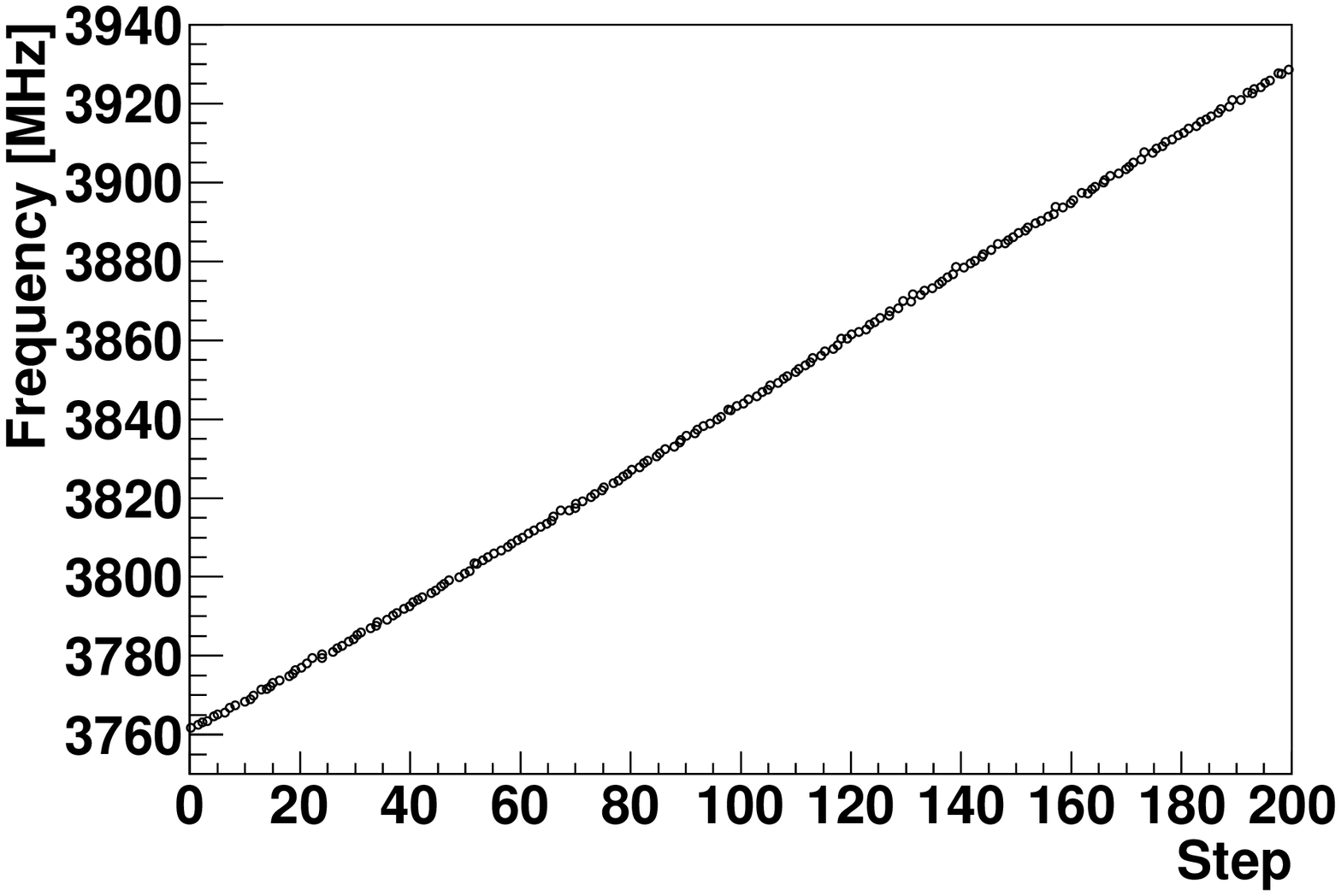}
\caption{(Color online.) Time durations for the tuning mechanism (top) and the TM$_{010}$ resonant frequency with step (bottom).
The red and blue filled histograms represent the time required for phase-matching and critical coupling, respectively, in seconds.
The former includes the time spent for target frequency shift.
The total time spent is represented by the empty black histogram.}
\label{fig:time_freq}
\end{figure}

This sequence of the tuning mechanism is repeated 200 times, with the target frequency being shifted by roughly 1\,MHz at every step.
This large frequency shift is chosen in order to encounter more occasions at which phase-matching is broken, so that the fine frequency tuning procedure is needed.
During the exercise, the time durations required to complete phase-matching and critical coupling are measured and the TM$_{010}$-like resonant frequency is recorded.
It is remarkable that the criteria for phase-matching (critical coupling) are satisfied more than 90\% (50\%) of the time, represented by the large red (blue) peak in Fig.~\ref{fig:time_freq} (a); for the rest 10\% (50\%) of the time, less than 2 (1) seconds are required to complete the process.
Note that the time duration for phase-matching includes the target frequency shift, which is about 0.5 second.
This indicates that the dead time owing to the procedure for tuning the frequency and coupling strength is insignificant in the data acquisition.
Good linear behaviour of the target frequency with step is also seen in Fig.~\ref{fig:time_freq} (b).

\section{Conclusion}
In this paper, we introduce a new design of a pizza-cylinder type multiple-cavity detector useful for detecting the high frequency regions of axion dark matter.
This design is characterized by multiple cells evenly divided by partitions and a narrow hollow gap in the middle of a cylindrical cavity.
We obtain the analytical solution and the quality factor, which are found to agree quite well with the numerical calculations.
One of the notable merits of this design is that it enhances the experimental sensitivity by making maximal use of the volume provided by the magnets.
Introduction of a narrow hollow in the middle of the cavity breaks the frequency degeneracy and eliminates the necessity of a power combiner, which simplifies the structure of the receiver chain.
Phase-matching is assured by vanishing of coupling strength for higher modes, which is visualized by disappearance of higher frequency peaks (circles) in the scattering parameter space (Smith chart).
Various simulation studies and an experimental demonstration verify that this design is promising for searching high mass regions in cavity-based axion experiments.

\section*{Acknowledgement}
This work is supported by IBS-R017-D1-2017-a00 / IBS-R017-Y1-2017-a00.
J.E.K. is also supported in part by the National Research Foundation grant funded by the Korean Government (NRF-2015R1D1A1A01058449).

\end{document}